\documentclass[14pt]{article}

\usepackage{arxiv}

\usepackage[utf8]{inputenc} 
\usepackage[T1]{fontenc}    
\usepackage{hyperref}       
\usepackage{url}            
\usepackage{booktabs}       
\usepackage{amsfonts}       
\usepackage{nicefrac}       
\usepackage{microtype}      
\usepackage{lipsum}		
\usepackage{graphicx}
\usepackage{natbib}
\usepackage{doi}

\usepackage{xcolor}
\usepackage{amsmath}

\makeatletter
\newcommand*{\rom}[1]{\expandafter\@slowromancap\romannumeral #1@}

\usepackage{xargs}
\usepackage{geometry}
\usepackage{fancyhdr} 
\usepackage{lastpage} 
\usepackage{extramarks} 
\usepackage[most]{tcolorbox} 
\usepackage{graphicx} 
\usepackage{xcolor} 

\usepackage[export]{adjustbox}
\usepackage[parfill]{parskip}
\usepackage{amssymb}
\newlength{\minuslength}
\usepackage{algorithm}
\usepackage{algorithmic}
\usepackage{caption} 
\usepackage{subcaption}
\usepackage{multirow}
\usepackage{setspace}
\setcitestyle{numbers}
\setcitestyle{square}
\setcitestyle{citesep={,}}

\usepackage{listings}
\lstdefinestyle{Rstyle}{
  language=R,
  basicstyle=\ttfamily\small,
  commentstyle=\color{gray},
  numbers=left,
  numberstyle=\tiny\color{gray},
  stepnumber=1,
  numbersep=5pt,
  backgroundcolor=\color{white},
  showspaces=false,
  showstringspaces=false,
  showtabs=false,
  frame=single,
  rulecolor=\color{black},
  tabsize=2,
  captionpos=b,
  breaklines=true,
  breakatwhitespace=false,
  keywordstyle=\color{blue},
  stringstyle=\color{orange}
}

\title{Non-stationary Bayesian Spatial Model for Disease Mapping based on Sub-regions}

\author{ \href{https://orcid.org/0000-0003-1587-3288}{\includegraphics[scale=0.06]{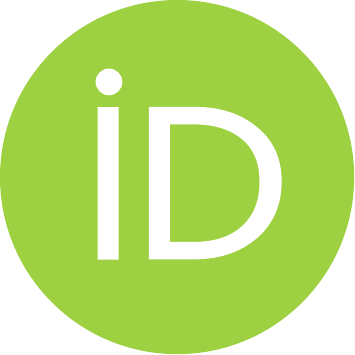}\hspace{1mm}Esmail Abdul-Fattah}\thanks{Corresponding Author} \\
	Statistics Program, CEMSE Division\\
	King Abdullah University of Science and Technology\\
	Thuwal, 23955, Makkah\\
	\texttt{esmail.abdulfattah@kaust.edu.sa} \\
        \And
        \href{https://orcid.org/0000-0002-7063-2615}Elias Krainski \\
        Statistics Program, CEMSE Division\\
	King Abdullah University of Science and Technology\\
	Thuwal, 23955, Makkah\\
	\texttt{elias.krainski@kaust.edu.sa } \\
        \And
	\href{https://orcid.org/0000-0002-4334-2057}Janet van Niekerk \\
	Statistics Program, CEMSE Division\\
	King Abdullah University of Science and Technology\\
	Thuwal, 23955, Makkah\\
	\texttt{janet.vanNiekerk@kaust.edu.sa}
        \And
	\href{https://orcid.org/0000-0002-0222-1881}H{\aa}vard Rue \\
	Statistics Program, CEMSE Division\\
	King Abdullah University of Science and Technology\\
	Thuwal, 23955, Makkah\\
	\texttt{haavard.rue@kaust.edu.sa} \\
}

\renewcommand{\shorttitle}{Non-stationary Bayesian Spatial Model for Disease Mapping based on Sub-regions}

\hypersetup{
pdftitle={A template for the arxiv style},
pdfsubject={q-bio.NC, q-bio.QM},
pdfauthor={David S.~Hippocampus, Elias D.~Striatum},
pdfkeywords={First keyword, Second keyword, More},
}

\begin{document}
\maketitle

\begin{abstract}

This paper aims to extend the Besag model, a widely used Bayesian spatial model in disease mapping, to a non-stationary spatial model for irregular lattice-type data. The goal is to improve the model's ability to capture complex spatial dependence patterns and increase interpretability. The proposed model uses multiple precision parameters, accounting for different intensities of spatial dependence in different sub-regions. We derive a joint penalized complexity prior for the flexible local precision parameters to prevent overfitting and ensure contraction to the stationary model at a user-defined rate. The proposed methodology can be used as a basis for the development of various other non-stationary effects over other domains such as time. An accompanying R package \texttt{fbesag} equips the reader with the necessary tools for immediate use and application. We illustrate the novelty of the proposal by modeling the risk of dengue in Brazil, where the stationary spatial assumption fails and interesting risk profiles are estimated when accounting for spatial non-stationary.

\end{abstract}


\section{Introduction}

In disease mapping models, the risk of disease occurrence is modeled on the geographical space on which the data generating mechanism is defined. Often the space consists of well-defined areas based on administrative boundaries from which the data is collected and aggregated. Bayesian hierarchical models can be used to explain and discover specific predictors that pertains to the acquiring of a disease ranging from individual covariates (like medical information, biographical information) to environmental covariates (climate, access to services such as healthcare, socio-economic factors). These models can then be used as part of a decision-making framework in the realm of public health to target interventions to the necessary areas and groups of people. Ideally, all covariates would be available but this is not often the case. Some covariates like access to healthcare can be very hard to measure on the individual level, leading to unexplained variation in the data. Random effects are designed to account for this unexplained variation in the form of temporal trends or spatial effects, amongst others. Disease mapping models usually contains at least one spatial random effect since the correlation between the risks of disease at nearby locations should be accounted for. This adds flexibility by accounting for the spatial correlation in ecological studies so that the relative risk in areas separated by a large space are less correlated in terms of disease risk than those in close vicinity. 

The data considered in disease mapping are usually aggregated by small areas, such as municipalities or counties. 
Therefore, the random effects are also usually specified at this areal level. The spatial model introduced by \cite{Besag1974SpatialIA} serves as the foundation for various models employed to analyze areal level data and capture spatial effects; Conditional Auto-Regression (CAR) model proposes a multivariate Gaussian distribution for the areal level spatial effects with a mean and covariance matrix which depends on the geographical structure.
The spatial effect is propositioned to be correlated to the spatial effects in nearby locations (i.e. neighbours) based on a neighbourhood set. The definition of the neighbourhood of a specific area is not fixed and various works on different definitions exists \citep{duncan2017}. Most commonly, a first order dependency is used whereby neighbours are defined through sharing at least one border. This neighbourhood has also been shown to be most applicable in practice \citep{duncan2017}. 
In this work we will thus restrict our approach to this first-order dependence in space and its intrinsic version. 

Let the Besag model for a random vector $\pmb x$ be an intrinsic Gaussian Markov random field (IGMRF) which is defined through a set of conditional distributions. Assume a first order neighbourhood structure and denote the neighbourhood of area $i$ as $i\sim j$, implying that area $j$ is a neighbour of area $i$.
Then, the conditional distribution for area $i$ (conditional on the other areas) is
\begin{equation}
    \pi(x_i|\pmb x_{-i}, \tau) \sim \mathcal{N} \Big(\frac{1}{n_i} \sum_{i \sim j} x_j, \frac{1}{n_i \tau} \Big).
    \label{condequ}
\end{equation}
This gives the conditional mean in \eqref{condequ} as the average of $\pmb x$ over the neighbors of $i$ and 
the conditional precision is $\tau n_i$, where $n_i$ is the number of the neighbors of area $i$. The smoothness of this model is governed by the common precision parameter $\tau$, resulting in equal intensity of the spatial effect across the entire domain.

The CAR model has been extended and reparameterized in various ways \citep{macnab2022bayesian}. Most notably, \citep{Besag1991BayesianIR} proposed a new spatial effect called the BYM (Besag-York-Mollie) model using the Besag model as part of a convolution. In the BYM model, the Besag structured effect is combined with an unstructured effect to account for within-area heterogeneity. Other spatial models include the Leroux model \citep{Leroux2000EstimationOD} which uses a combination of the precision matrices from each model component in the BYM model, and the Dean model \citep{Dean2001DetectingIB} proposed a new parametrization for the (same) model in \citep{Besag1991BayesianIR} (i.e. a new way to combine the random effects in \citep{Besag1991BayesianIR}), while \citep{Riebler2016AnIB} reparametrized it for interpretability. The development of these spatial models inclusively involves the presence of structured and unstructured spatial effects in the same model, the interpretability of the hyperpriors, the scaling of the adjacency spatial structure, and the choice of prior distributions for the model parameters. Another approach is based on basis functions and considering the Besag model, or its variations, as priors for the coefficients \citep{Lee2022ABM}. A comparison between the CAR model and a P-spline model can be found in the work of \citep{goicoa2012comparing}.

The parameter $\tau$ in Eq. (1) is assumed to be constant over the domain, i. e., the same for all the areas.
This assumption have not been challenged extensively in the literature, yet this assumption may be too restrictive, especially for a large spatial domain with many small areas.
For a large number of areas \citep{OrozcoAcosta2021ScalableBM} proposed a
``divide and conquer" approach dividing the spatial region into sub-regions, defined as groups of contiguous small areas. Multiple fits are then averaged to arrive at a final model where the intensity of the spatial effect is different for the different sub-regions. Besides the computational motivation evident in the "divide and conquer" approach, it may be reasonable to assume that the properties of different sub-regions (groups of small areas) can be different. 
For instance, Germany could be divided into four sub-regions each one having $p_i$, $i = 1, \ldots, 4$, small areas, 
as illustrated in Figure \ref{fig:partgerm}.
Within each sub-region $i$ the $p_i$ small areas could have a precision parameter $\tau_i$,
allowing the variability of the spatial effects to vary across sub-regions. This results in a non-stationary spatial model for areal data. The choice of sub-regions is not a primary objective of this paper, there are several possible ways to divide a spatial region based on factors such as geographical, climate, administrative, historic, and economic features. 
One could expect higher variability in the disease risk in a sub-region of the domain than in others where it may be less pronounced.
Also, vector-borne illnesses are more prevalent in sub-regions with warmer climates, as the disease-carrying organisms are more active in these environments 
and it may lead to a different degree of local variability in the disease risk.
Furthermore, a spatial region can be divided into sub-regions based on the proximity of the small areas to each other and the presence of largely uninhabited areas in each sub-region.

As such, we propose an extension of the Besag model 
where the conditional distributions assume the precision parameter to vary between the sub-regions. We propose to form these sub-regions as groups of contiguous small areas similar to \citep{OrozcoAcosta2021ScalableBM}. 
The derived precision matrix of the spatial effect allows the intensity of the spatial dependence to change over 
the sub-regions while preserving the contiguity at the edges of the sub-regions. This proposal allows the inference to be done in one model fit while capturing a more complex non-stationary dependency structure. The new non-stationary flexible Besag (\texttt{fbesag}) model is proposed in Section \ref{sec3pbesag}. We derive the joint penalizing complexity prior for the parameters of the new non-stationary flexible Besag (\texttt{fbesag}) model to prevent overfitting and ensure contraction to the stationary model if the data necessitates it.  
In Section \ref{sec:App} we demonstrate the applicability of the proposal by modeling the incidence of dengue in Brazil using the Integrated Nested Laplace Approximations (INLA).
Finally, we summarize the conclusions in Section~\ref{secconclusions}.

\section{Dengue in Brazil}\label{sec:motivation}
Dengue is a vector-borne viral infection caused by the dengue virus. It is mostly found in tropical and sub-tropical climates where the infected mosquitoes naturally occur and breed. Dengue can cause sever illness or death and has no specific cure. Early detection and vector control are vital for lower fatality rates of severed dengue. Around the world, the incidence of dengue has increased severely from around 500 000 reported cases in 2000 to 5.2 million in 2019. The disease is endemic to more than 100 countries globally with Brazil being one of the most affected countries \citep{bhatt2013global}.  

Disease mapping plays an important role in understanding the evolution of a disease and informing intervention policies necessary for the protection of public health. As such, the unexplained spatial variability of a disease provides the relevant stakeholders with interesting patterns hidden by the considered predictors such as socio-economic indicators, administrative policies at a region-level, region-level geographical characteristics etc. Many studies have shown associations between climate factors and dengue risk, and \cite{Lowe2021CombinedEO} presents a better understanding of the effects of extreme droughts and wet conditions on the disease. \cite{Lowe2021CombinedEO} used the Besag model to analyze the effects of hydrometeorological hazards on dengue risk in Brazil. Further research is necessary to better understand the dengue virus as the disease poses a significant threat. The Besag spatial field is commonly used in disease mapping models and have also been used in the modeling of dengue. However, when using the Besag model we are assuming that the spatial field is stationary, i.e. that the variation around the mean is constant for each location in space. This assumption can be too strict especially in countries with many administrative areas or towns.   

\begin{figure}[hb!]
    \centering
    \begin{subfigure}{.5\textwidth}
      \centering
      \includegraphics[width=\linewidth]{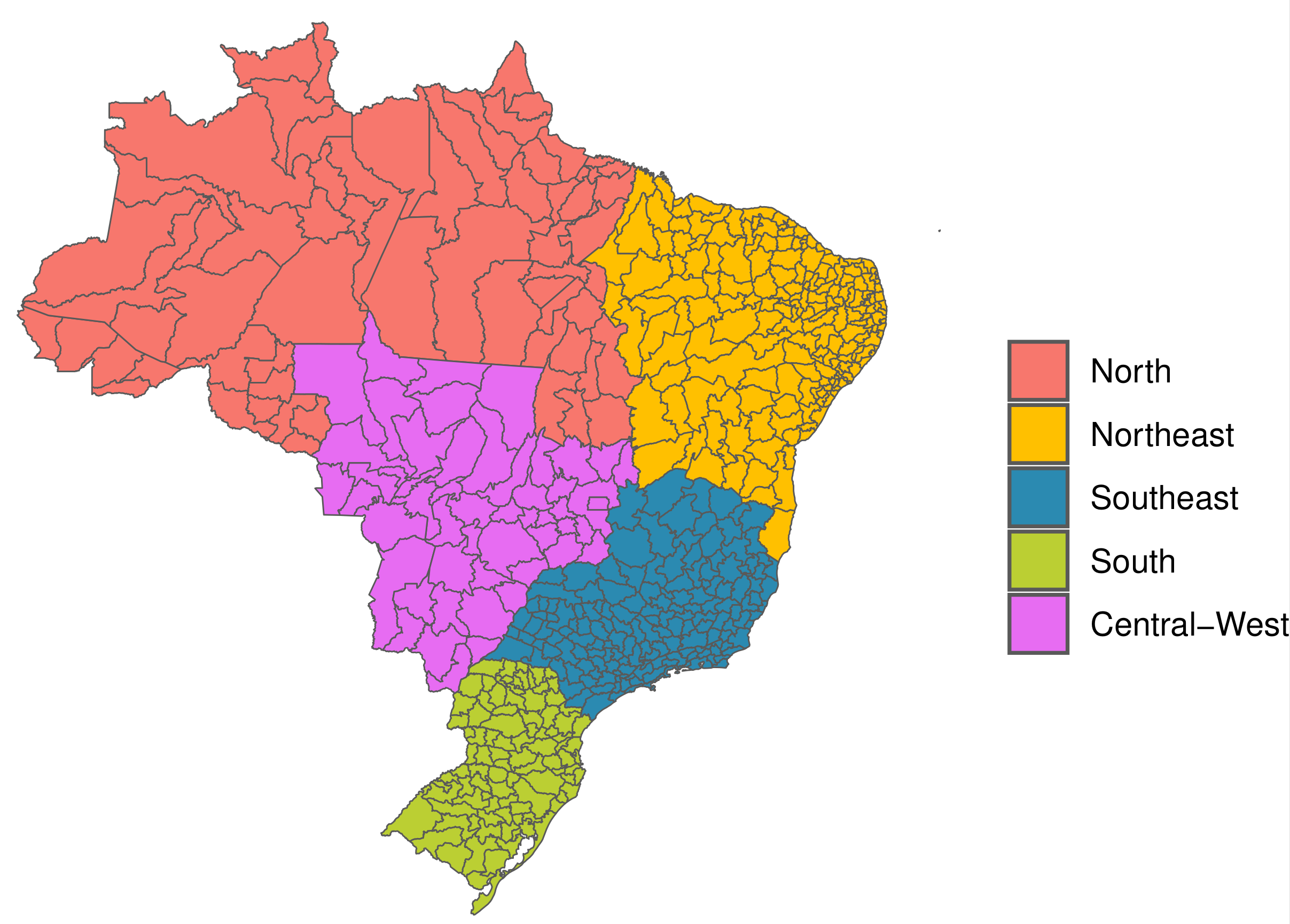}
    \end{subfigure}%
    \begin{subfigure}{.5\textwidth}
      \centering
      \includegraphics[width=\linewidth]{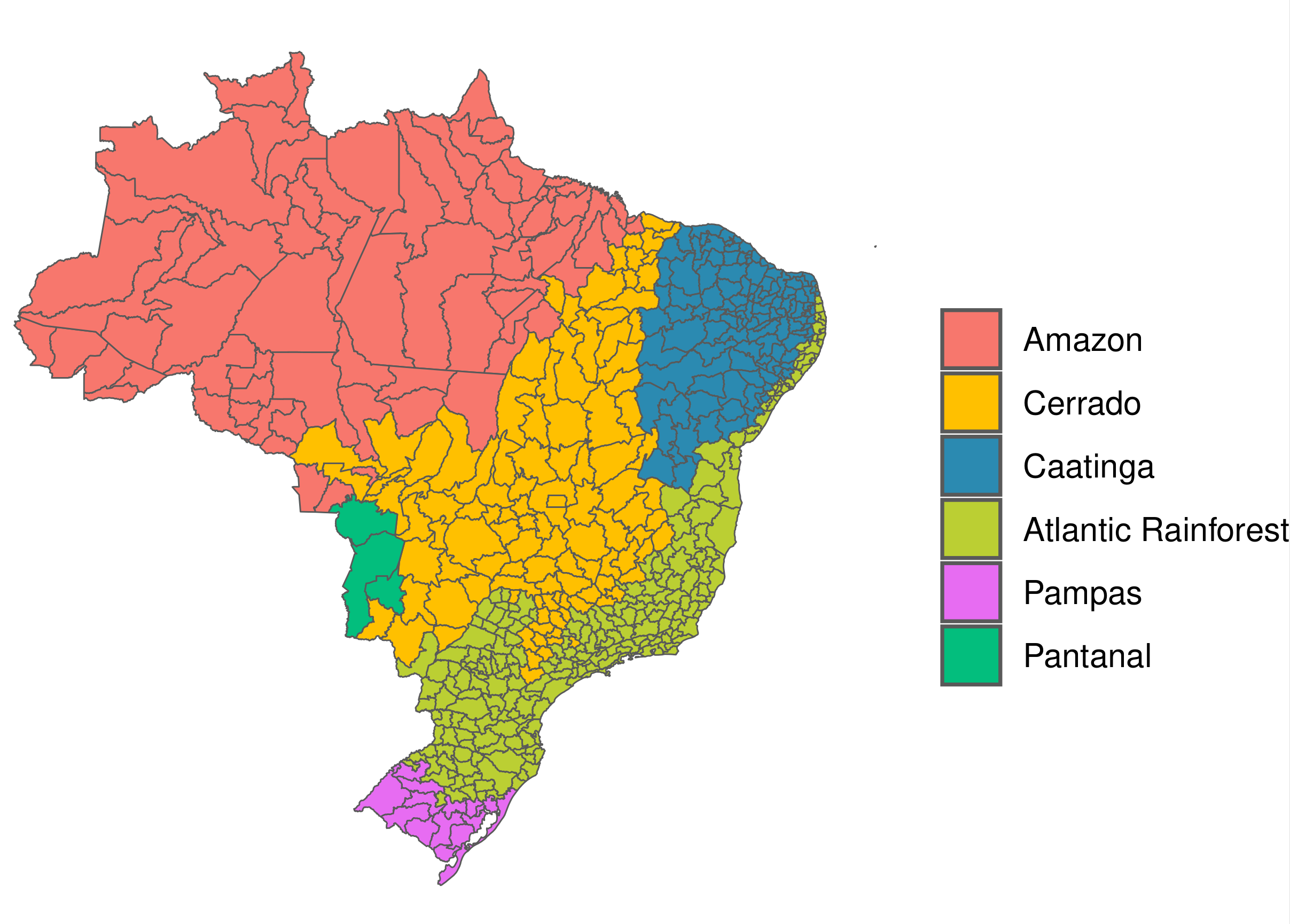}
    \end{subfigure}
    \caption{Subgroups of regions in Brazil based on administrative characteristics (left) and terrestrial biomes (right)}
   \label{fig:brazilsub}
\end{figure}

Brazil is divided into 558 microregions. Assuming that the spatial variation has the same intensity in each of these 558 microregions seems a bit unrealistic. We could advocate that in smaller subgroups of microregions the spatial field has the same intensity, but the intensity can be different between subgroups. Allowing for different intensities across subgroups then result in a non-stationary spatial field. This leads us to ponder the formation of these subgroups of the microregions. For Brazil, we can group the microregions in various ways of which we illustrate two options in Figure \ref{fig:brazilsub}. The choice of the subgroups should be dictated by the purpose of the analysis in a sense that if we want to investigate the risk of dengue for different administrative regimes then we would consider the subgroups based on that, otherwise if we are interested to discover spatial patters across the different biomes then we would use those characteristics to form subgroups. The sub-regions in Figure \ref{fig:brazilsub} (left) are defined based on geographical, political, administrative, and economic differences. In Figure \ref{fig:brazilsub} (right) we consider the six Brazilian terrestrial biomes: Amazon Rainforest, Atlantic Forest, Caatinga, Cerrado, Pampas, and Pantanal. Variations in climatic conditions, vegetation, and wildlife characterize these sub-regions. 

The possible non-stationarity of the spatial field should be accounted for in the disease mapping framework to produce accurate predictions of risk. In this work we propose an extension of the Besag model that incorporates the non-stationarity while also being able to contract to stationarity with the use of a contraction prior for the flexibility parameters. Without the contraction prior, the risk of overfitting a stationary field with a non-stationary field is high and this makes the proposal unattractive in practice. This contraction property is essential for practitioners since it is not clear how to decide on a stationary or non-stationary model for irregular lattice data, like in disease mapping. We aim to provide a new non-stationary spatial model for irregular lattice data that can be used i) to fit a model to a non-stationary spatial field and ii) as an investigative tool into the non-stationarity of a field based on the contraction property. This new framework can provide deeper and more accurate insights into disease evolution that can be used to improve general public health outcomes, like for dengue in Brazil.

\section{Non-Stationary flexible Besag Model (\texttt{fbesag}) based on sub-regions} \label{sec3pbesag}
The \texttt{fbesag} model is proposed for $\pmb {x}$, as a non-stationary alternative to the stationary Besag model for a spatial domain consisting of $N$ small areas and $P$ subregions, $P\leq N$. 
The joint density function of $\pmb x$ with precision parameters $\tau_1, \tau_2, ...,\tau_P$ is defined as
    \begin{equation}
            \pi(\pmb x|\tau_{1}, \ldots, \tau_{P}) \propto \exp\Big(-\dfrac{1}{4} \sum_
            {\substack{i\text{ in sub-region } k \\ j\text{ in sub-region } l \\ i \sim j \\ i > j }} (\tau_{l} + \tau_{k} )(x_i - x_{j})^2 \Big), \quad  k, l = 1, \ldots, P,
            \label{eq::besagtype1}
        \end{equation}
    \noindent with conditional densities
       \begin{equation*}
        x_i |\pmb x_{-i}, \tau_{1}, \ldots, \tau_{P} \sim \mathcal{N} \Big(\dfrac{1}{2}\displaystyle \sum_{\substack{i\text{ in sub-region } k \\ j\text{ in sub-region } l \\ i \sim j}} (\tau_{l} + \tau_{k})\tau_{x_i}^{-1} x_j, \tau_{x_i}^{-1}\Big),
    \end{equation*} and
    \begin{equation*}
        \tau_{x_i} = \dfrac{1}{2}\Big(n_{i} \tau_{k}  + \sum_{l} n_{il} \tau_{l}\Big).
    \end{equation*}
\noindent Note that $\tau_{x_i}$ is the precision of county $i$, $i \sim j$ represents the neighbors of $i$, $i \neq j$, $n_i$ is the number of neighbors of area $i$, $n_{il}$ is the number of neighbors of area $i$ in sub-region $l$, and $\tau_l$ is the precision parameter of sub-region $l$. Since all the conditional density functions are univariate Gaussian, $\pmb x$ will follow a multivariate Gaussian distribution with a precision matrix (inverse covariance matrix) derived from the neighbourhood structure. We denote this model as $$\pmb x|\tau_1, \tau_2, ..., \tau_P\sim \texttt{fbesag}(\tau_1, \tau_2, ..., \tau_P).$$

Consider the stationary case where all the precision parameters are equal, i.e. $\tau_1=\tau_2 = ... = \tau_P = \tau$, then \eqref{eq::besagtype1} reduces to
 \begin{equation*}
            \pi(\pmb x|\tau) \propto \exp\Big(-\dfrac{1}{4} \sum_
            {i \sim j } 2\tau (x_i - x_{j})^2 \Big) = \exp\Big(-\dfrac{1}{2} \sum_
            {i \sim j } \tau (x_i - x_{j})^2 \Big),
        \end{equation*}
hence the \texttt{fbesag} model reduces to the Besag model for $\tau_1=\tau_2 = ... = \tau_P = \tau$.

\begin{figure}
\centering
\begin{subfigure}{.35\textwidth}
  \centering
  \includegraphics[width=\linewidth]{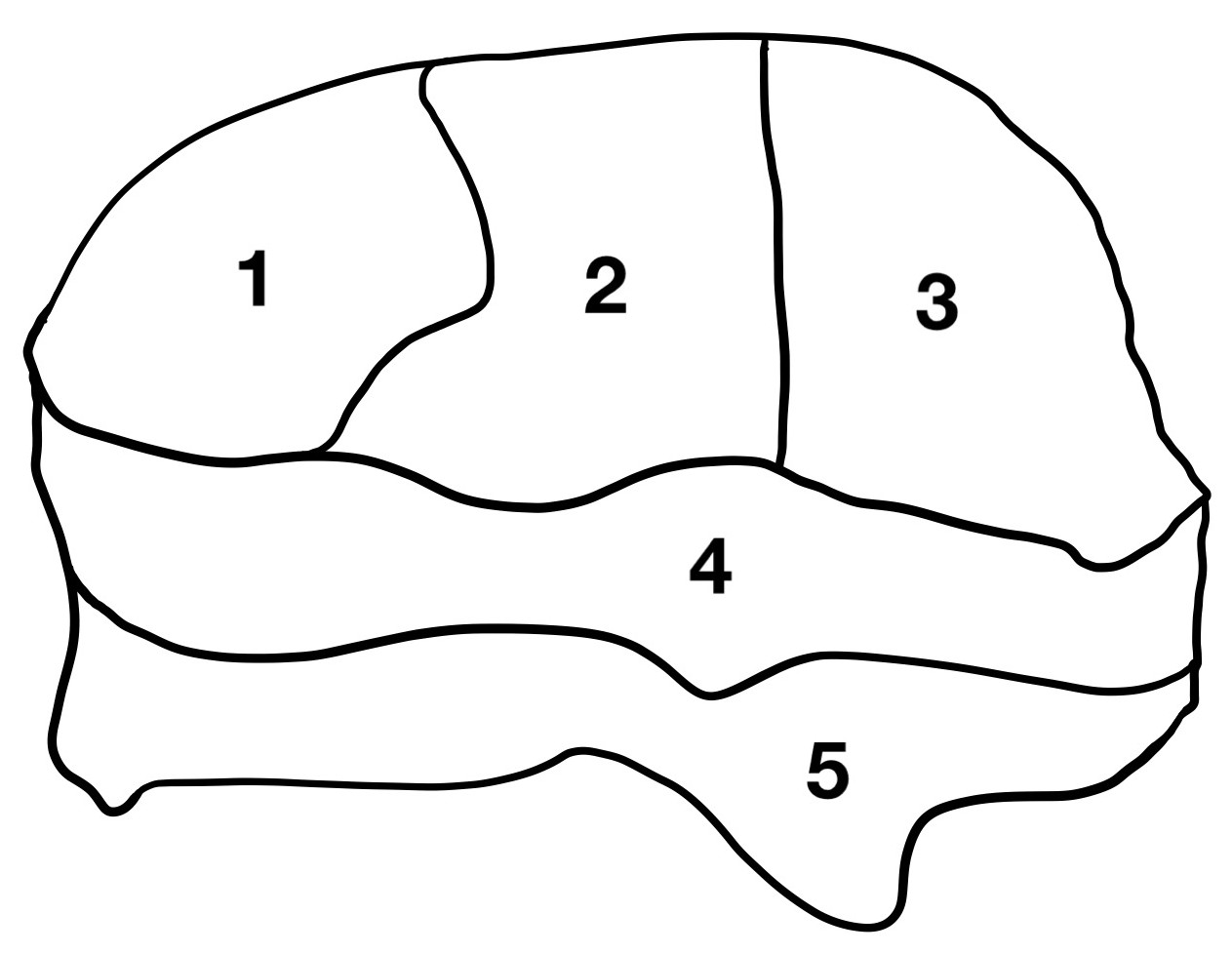}
\end{subfigure}%
\begin{subfigure}{.35\textwidth}
  \centering
  \includegraphics[width=\linewidth]{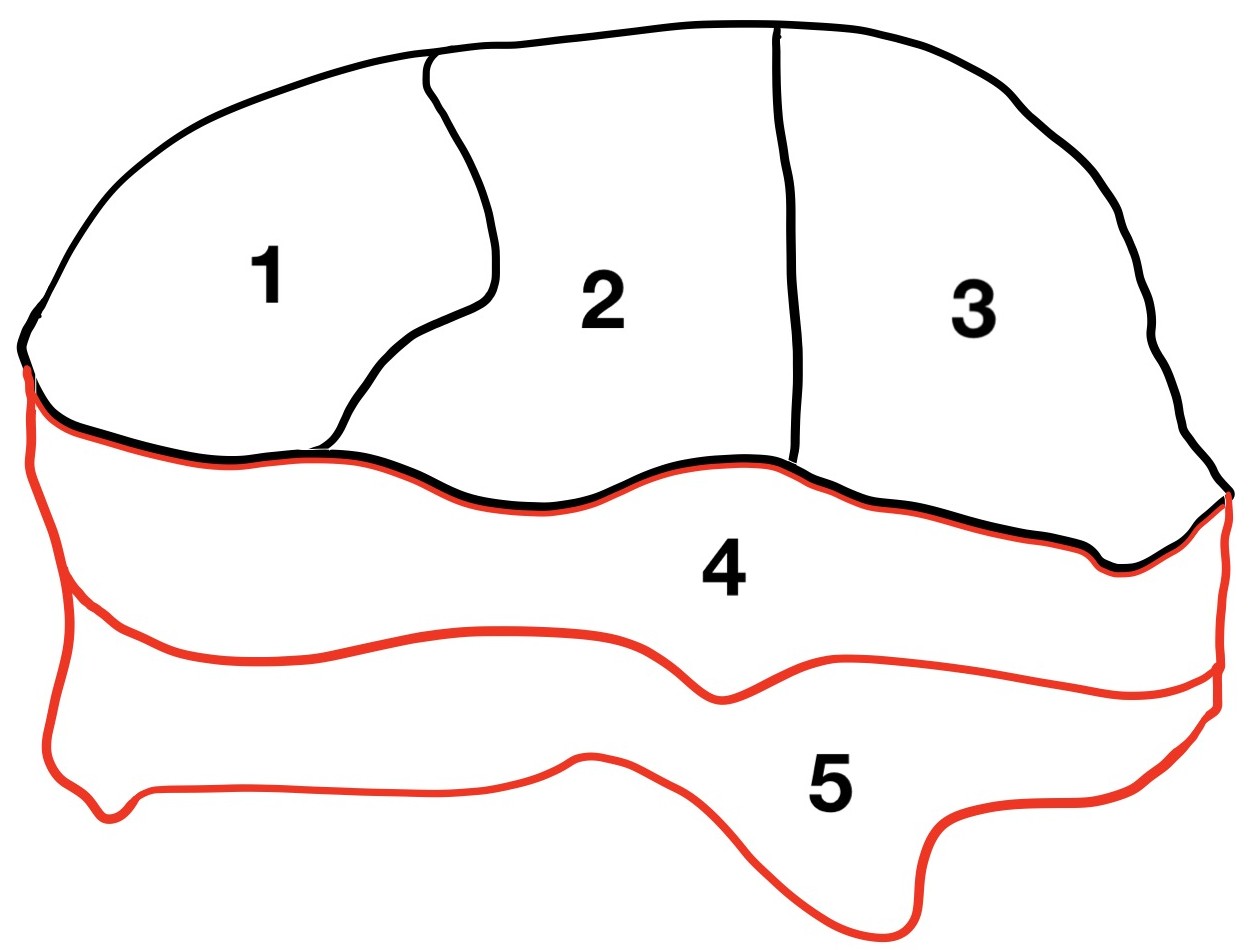}
\end{subfigure}
\caption{Illustration of \texttt{fbesag} models with one sub-region and two sub-regions}
     \label{fig:twopartitionsillu}
\end{figure}

\subsection{Example} Consider the case of five areas in the spatial domain. Suppose we are considering one or two subregions. Figure \ref{fig:twopartitionsillu} presents the domain with one (left) or two (right) sub-regions. The precision parameters for the case with two sub-regions $c_1 = \{1, 2, 3\}$ (black part) and $c_2 = \{4, 5\}$ (red part) are $\tau_{1}$ and $\tau_{2}$, respectively, while the stationary model with one subregion has one precision parameter $\tau_{1}$. From (\ref{eq::besagtype1}) we can construct the precision matrix for the one sub-region (stationary model) as,
\[ \pmb Q_{S} =  \left( \begin{array}{ccccc}
2\tau_{1} & -\tau_{1} & 0 & -\tau_{1} & 0 \\
-\tau_{1} & 3\tau_{1} & -\tau_{1} & -\tau_{1} & 0 \\
0 & -\tau_{1} & 2\tau_{1} & -\tau_{1} & 0 \\
-\tau_{1} & -\tau_{1} & -\tau_{1} & 4\tau_{1} & -\tau_{1} \\
0 & 0 & 0 & -\tau_{1} & \tau_{1}
\end{array} \right),
\]
and the precision matrix for the non-stationary model based on two sub-regions is constructed as,
\[
\pmb Q_{NS} = \left( \begin{array}{ccccc}
\tfrac{3}{2}\tau_{1} \textcolor{red}{+\tfrac{1}{2}\tau_{2}} & -\tau_{1} &  0 & -\tfrac{1}{2}\tau_{1} \textcolor{red}{-\tfrac{1}{2}\tau_{2}} & \textcolor{red}{0} \\
-\tau_{1} & \tfrac{5}{2}\tau_{1} \textcolor{red}{+ \tfrac{1}{2}\tau_{2}} & -\tau_{1} & -\tfrac{1}{2}\tau_{1} \textcolor{red}{-\tfrac{1}{2}\tau_{2}} & \textcolor{red}{0} \\
0 & -\tau_{1} & \tfrac{3}{2}\tau_{1} \textcolor{red}{+ \tfrac{1}{2}\tau_{2}} & -\tfrac{1}{2}\tau_{1} \textcolor{red}{-\tfrac{1}{2}\tau_{2}} & \textcolor{red}{0} \\
-\tfrac{1}{2}\tau_{1} \textcolor{red}{-\tfrac{1}{2}\tau_{2}} & -\tfrac{1}{2}\tau_{1} \textcolor{red}{-\tfrac{1}{2}\tau_{2}} & -\tfrac{1}{2}\tau_{1} \textcolor{red}{-\tfrac{1}{2}\tau_{2}} & \tfrac{3}{2}\tau_{1} \textcolor{red}{+ \tfrac{5}{2}\tau_{2}} & \textcolor{red}{-\tau_{2}} \\
\textcolor{red}{0} & \textcolor{red}{0} & \textcolor{red}{0} & \textcolor{red}{-\tau_{2}} & \textcolor{red}{\tau_{2}}
\end{array} \right)
\]
where $\pmb Q_{S}$ and $\pmb Q_{NS}$ are the precision matrices of the \texttt{fbesag} model having one or two sub-regions, respectively.
In the case of two sub-regions, area 5 is located within sub-region 2 and all of its neighboring small areas are also in sub-region 2, hence area 5 is assigned the same precision parameter as its corresponding sub-region. However, area 4 has neighbors from both sub-region 1 and sub-region 2, so it is assigned a mixture of precision parameters from the two sub-regions.
Note that if the two sub-regions have the same precision i.e. $\tau_1 = \tau_2$, then the first term in $\pmb Q_{NS}$ will be $\tfrac{3}{2}\tau_{1} +\tfrac{1}{2}\tau_{2} = 2 \tau_1 = 2 \tau_2$, and subsequently $Q_{NS}$ will simplify to $Q_S$, which is the precision matrix of the stationary Besag model from \cite{Besag1974SpatialIA}. 

\subsection{Joint PC prior for the local precision parameters} Flexible models which are based on simpler models, like the \texttt{fbesag} model, could overfit the data due to the additional flexibility. To prevent this, we propose a joint penalizing complexity (PC) prior for the precision parameters. This prior is constructed to contract to the stationary model (with a single precision parameter) if the data does not provide sufficient evidence for the non-stationary model. Using a PC prior provides the additional benefit of using the \texttt{fbesag} model as an investigative tool in determining the stationarity of a spatial field. If we are at risk of overfitting then we cannot confidently conclude non-stationarity based on a fit of the \texttt{fbesag} model, since the non-stationary model will almost surely provide a spatial field that is non-stationary. Hence, the invoking of the PC prior framework is to ensure the appropriate contraction to stationarity.  
We follow the penalized complexity prior \citep{Simpson2014PenalisingMC} framework to derive a joint contraction prior for the local precision parameters in the \texttt{fbesag} model.  

As a stationary model we consider the Besag model with precision parameter $\tau$.
 Now, consider the \texttt{fbesag} model with $P$ subregions and corresponding local precision parameters $\tau_1, \ldots, \tau_P$, as the flexible counterpart. Define $\tau_j = \tau e^{\gamma_j}$, such that if $\gamma_j = 0$ then the local precision $\tau_j$ is equal to $\tau$. Define $\pmb \tau = (\tau e^{\gamma_1}$, \ldots, $\tau e^{\gamma_P})$. The joint PC prior for $\pmb\theta = \log\pmb\tau$ can be derived as a convolution of the PC prior for $\tau$ from the Besag model and an i.i.d. prior for the elements of $\pmb\gamma$ such that $\gamma_j\sim \mathcal{N}(0, \sigma^2_\gamma)$, as follows
 \begin{equation}
         \pi(\pmb \theta) = 2^{-(P + 2)/2} \pi^{-P/2} \lambda \sigma^{-P}  \exp \Big(-\frac{1}{2} (\pmb \theta- \pmb 1 \overline{\theta})^T \tilde{\pmb \Sigma}^{-1} (\pmb \theta- \overline{\theta} \pmb 1) - \overline{\theta}/2 - \lambda e^{-\overline{\theta}/2}\Big),
         \label{Jointprior}
    \end{equation}
\noindent where $\overline{\theta}$ is the mean of $\pmb \theta$, $\tilde{\pmb \Sigma} = \sigma^2_{\gamma} (\pmb I - \frac{1}{P} \pmb 1_{P \times P})$, $\lambda = -\dfrac{\ln \alpha}{u}$ and the variability of $\gamma_{j}$, for $j = 1, 2, ..., P$, is determined by the flexibility parameter $\sigma_{\gamma}$. Note that $\lambda$ satisfies
\begin{equation}
   \text{prob}(\tau^{-1/2} > u) = \alpha, ~~ u > 0, ~~ 0 < \alpha < 1.
\end{equation}

The parameters $\lambda$ and $\sigma_{\gamma}$ can be selected by the user based on some prior knowledge of probable regions for $\pmb \tau$.  This joint prior gives proper contraction to $\pmb \gamma = 0$ and decays with increasing complexity so that the simplest model is favored unless there is evidence for a more flexible model. 

A detailed derivation of the joint PC prior is presented in Appendix \ref{appendix::pbesagPC}.
The onus is on the practitioner to define a set of sub-regions for the spatial region. Some insights into this can be found in \cite{OrozcoAcosta2021ScalableBM}.
The proposed model thus incorporates the spatial strength of small areas at the boundary of the sub-regions depending on the mix of precision parameters from the neighboring sub-regions. 
When the number of small areas is large and the spatial region can be split into sub-regions, the proposed model might discover interesting spatial patterns. 

\section{Simulation examples}\label{sec:App}

\begin{figure}
\centering
\begin{subfigure}{.33\textwidth}
  \centering
  \includegraphics[width=\linewidth]{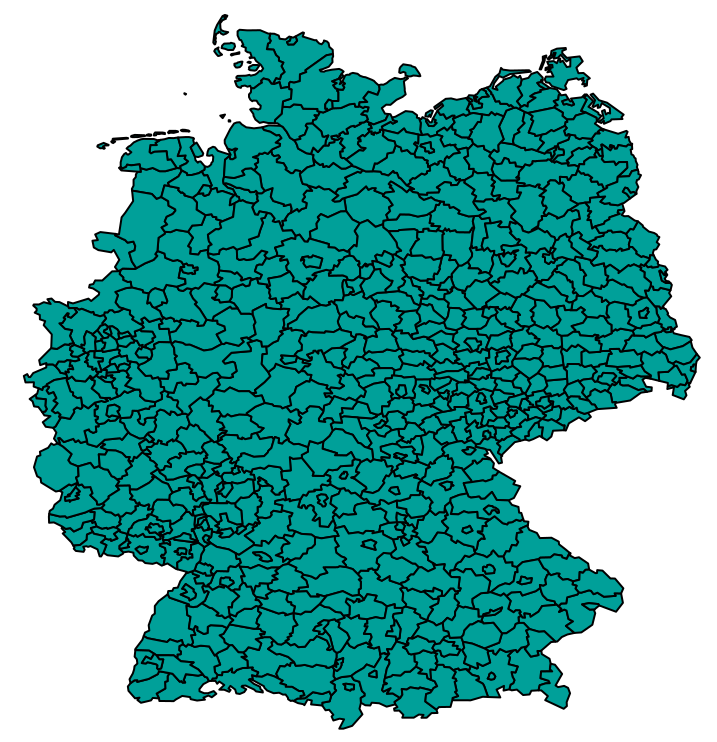}
  \caption{stationary}
\end{subfigure}%
\begin{subfigure}{.33\textwidth}
  \centering
  \includegraphics[width=\linewidth]{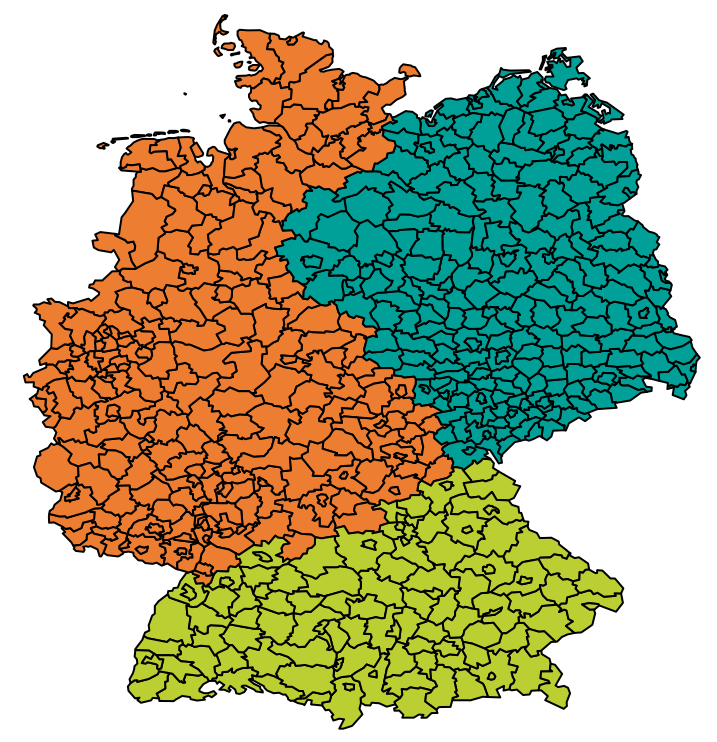}
  \caption{3 sub-regions}
\end{subfigure} %
\begin{subfigure}{.33\textwidth}
  \centering
  \includegraphics[width=\linewidth]{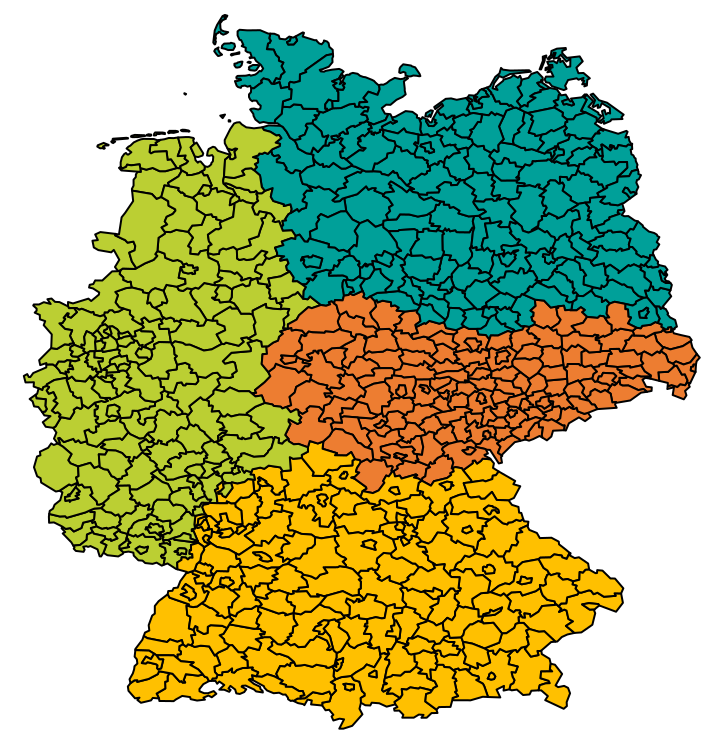}
  \caption{4 sub-regions (A)}
\end{subfigure} \\
\begin{subfigure}{.33\textwidth}
  \centering
  \includegraphics[width=\linewidth]{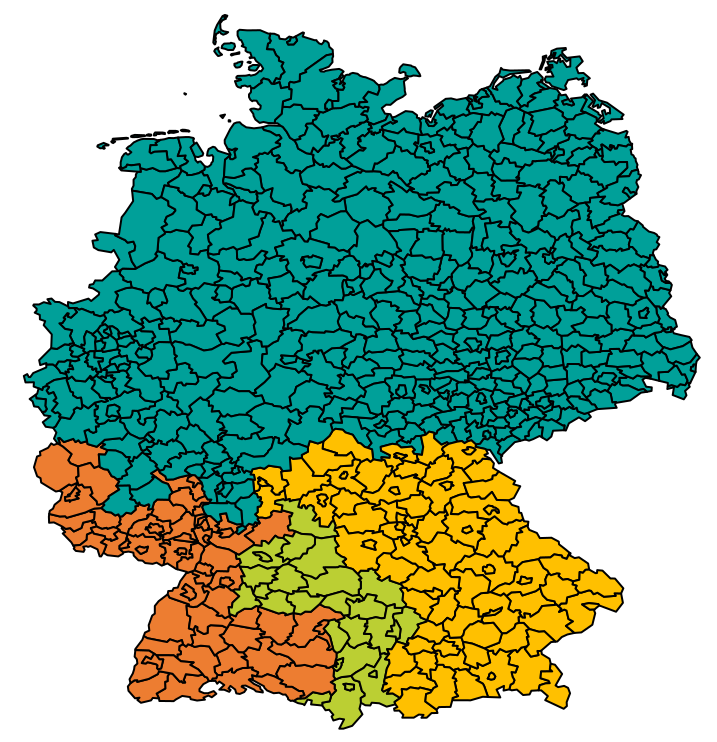}
  \caption{4 sub-regions (B)}
\end{subfigure}%
\begin{subfigure}{.33\textwidth}
  \centering
  \includegraphics[width=\linewidth]{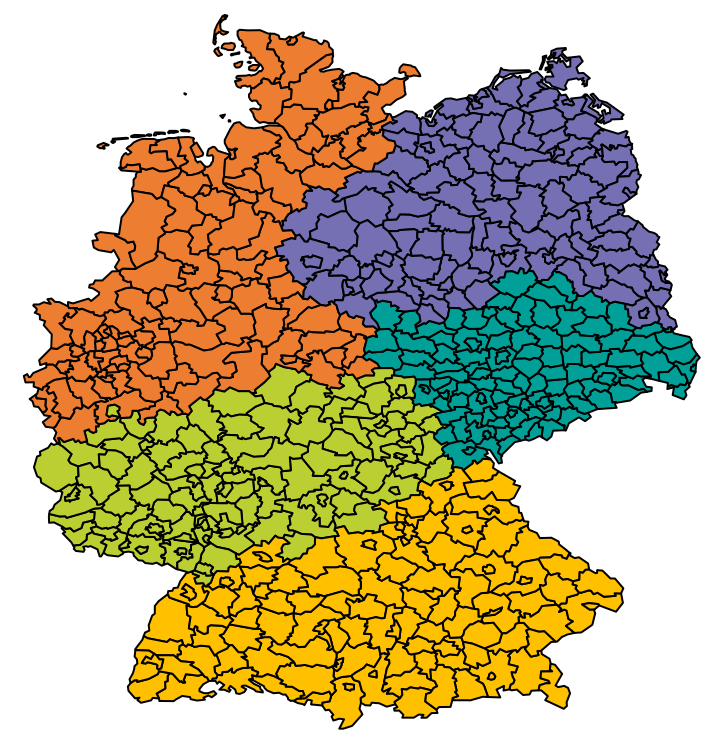}
  \caption{5 sub-regions}
\end{subfigure}%
\begin{subfigure}{.33\textwidth}
  \centering
  \includegraphics[width=\linewidth]{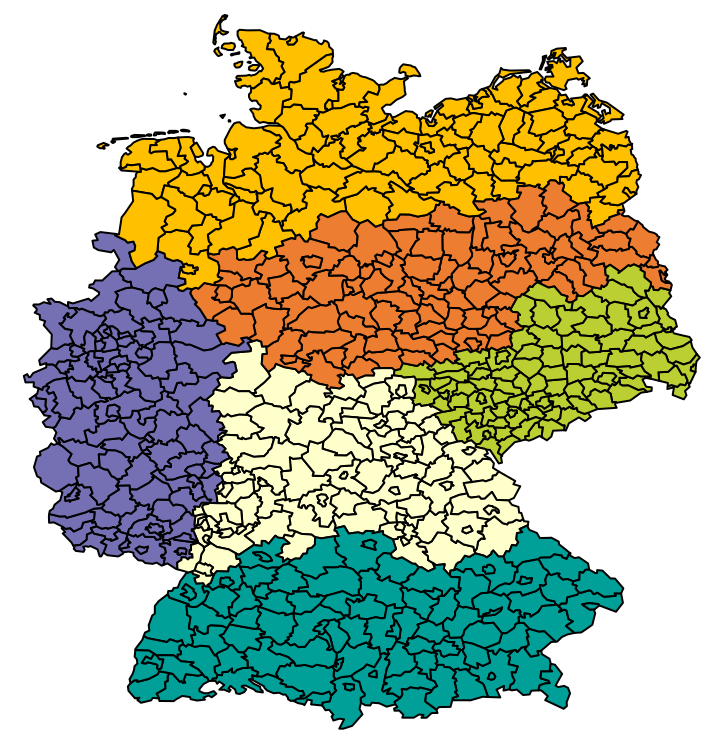}
  \caption{6 sub-regions}
\end{subfigure}
\caption{Illustration of different sub-regions used for \texttt{fbesag} models of Germany}
     \label{fig:partgerm}
\end{figure}

In this section, we simulate different data to which we apply the flexible Besag model and we analyze the results. For this simulation study we considered the domain of Germany which is divided into 544 German districts.
We defined five different configurations of sub-regions as shown in Figure \ref{fig:partgerm}, from one sub-region to six sub-regions.

In this simulation study we aim to show three properties of the model. Firstly, model's ability to uncover the true model and the most likely sub-regions from a set of possible sub-regions, then we explore the effect of the choice of the flexibility parameter $\sigma_\gamma$ and lastly we show how the local precision parameters contract to a global value when the true model is the stationery Besag model.  

\subsection{Recovering the true sub-regions and local precision parameters}
We simulate Poisson counts using the flexible Besag model with 4 sub-regions (A) as in Figure \ref{fig:partgerm}(c), using the following regression model,
\begin{equation}
    \pmb y \sim \text{Poisson}(\exp( \pmb 2 + \pmb \alpha)) \;,\label{eq:poisson_sim}
\end{equation}
with $\pmb \alpha|\tau_1, \tau_2, \tau_3, \tau_4\sim \texttt{fbesag}(\tau_1, \tau_2,  \tau_3, \tau_4).$
We fit the simulated data with a stationary Besag model, an \texttt{fbesag} model with 4 sub-regions (A) and an \texttt{fbesag} model with 4 sub-regions (B). The true values of the local precision parameters are calculated from six different values of $\tau$: $-2, -1, 0, 1, 2, $ and $ 3$, and for each one we simulate $\pmb\gamma \sim \mathcal{N}(0, 0.2^2\pmb I)$ and calculate
\begin{equation}
        \pmb \tau = \tau e^{\pmb \gamma}.
\end{equation}
Thus ($\tau_1$, $\tau_2$, $\tau_3$, $\tau_4$) vary around $\tau$. For each one of the six sets of fixed $\pmb\tau$ we generate $100$ realizations and use INLA \citep{Martins2013BayesianCW} to fit the model to the data. We assign $\lambda = 11.51$, i.e. $\text{prob}(\tau<1) = 0.00001$ \citep{Simpson2014PenalisingMC} and take $\sigma_\gamma = 0.2$ for the parameters of the hyperprior \eqref{Jointprior}. 

\begin{figure}
\centering
\begin{subfigure}{.5\textwidth}
  \centering
  \includegraphics[width=\linewidth]{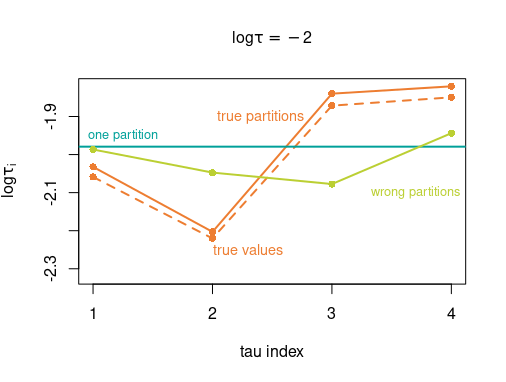}
\end{subfigure}%
\begin{subfigure}{.5\textwidth}
  \centering
  \includegraphics[width=\linewidth]{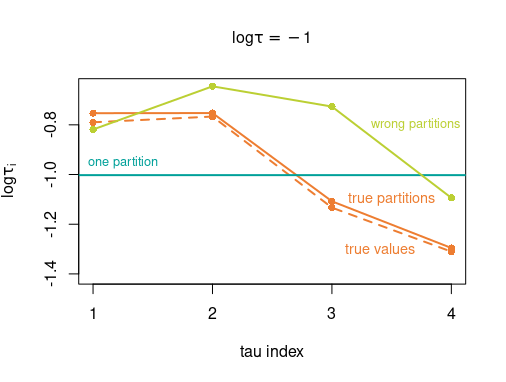}
\end{subfigure} \\
\begin{subfigure}{.5\textwidth}
  \centering
  \includegraphics[width=\linewidth]{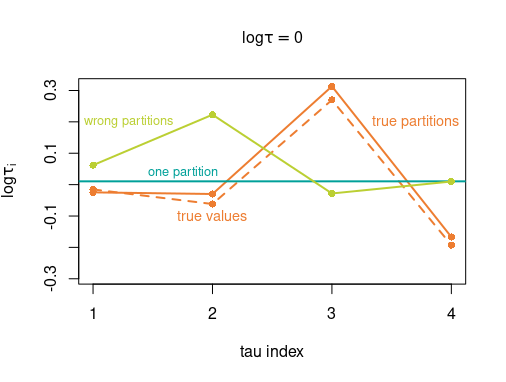}
\end{subfigure}%
\begin{subfigure}{.5\textwidth}
  \centering
  \includegraphics[width=\linewidth]{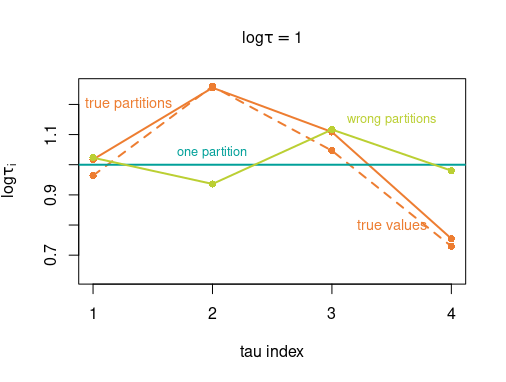}
\end{subfigure}\\
\begin{subfigure}{.5\textwidth}
  \centering
  \includegraphics[width=\linewidth]{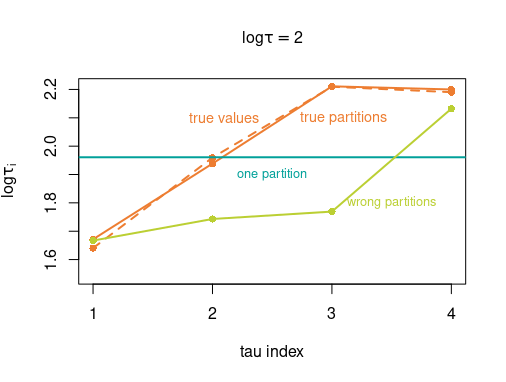}
\end{subfigure}%
\begin{subfigure}{.5\textwidth}
  \centering
  \includegraphics[width=\linewidth]{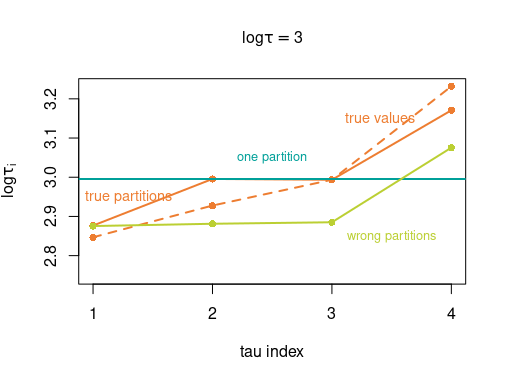}
\end{subfigure}
\caption{Results of log precision values after simulating data for Poisson counts using the flexible Besag model (4 sub-regions (A)) and comparison with two wrong models: one sub-region and four wrong sub-regions}
     \label{fig:wrongright}
\end{figure}

In Figure \ref{fig:wrongright}, we compare the true values $\log \pmb \tau_{\text{true values}}$ with $\log \tau_{\text{stationary}}$, $\log \pmb \tau_{\text{4 sub-regions (A)}}$, and $\log \pmb \tau_{\text{4 sub-regions (B)}}$. 
The results show that the true model is recovered using 4 sub-regions (A) for each set of true values based on the true value of $\tau$. The mean of the $\log \pmb \tau_{\text{4 sub-regions (A)}}$ is almost equal to the $\log \tau_{\text{stationary}}$ we obtain from the stationary Besag model. Unlike the \texttt{fbesag} model with 4 sub-regions (B) and the stationary Besag, the \texttt{fbesag} model with 4 sub-regions (A) captures the true parameter values with a small uncertainty. Selecting the wrong sub-regions is not necessarily detrimental, as we may still achieve improved results when compared to the stationary Besag model; see Table \ref{DIClogmlik:fbesag} for a comparison of the DIC (Deviance Information Criterion) and logML (log marginal likelihood) measures for the three models: the stationary Besag model, the \texttt{fbesag} model with 4 sub-regions (A), and the \texttt{fbesag} model with 4 sub-regions (B). 

\begin{table}
\centering
\renewcommand{\arraystretch}{1.3}
\begin{tabular}{c|cll|cll}
\multicolumn{1}{l|}{\textbf{}}           & \multicolumn{3}{c|}{\textbf{Sub-regions}}                                                             & \multicolumn{3}{c}{\textbf{Sub-regions}}                                                             \\ \cline{2-7} 
\textbf{}                                & \multicolumn{1}{c|}{\textbf{one region}} & \multicolumn{1}{c|}{\textbf{4-A}} & \multicolumn{1}{c|}{\textbf{4-B}} & \multicolumn{1}{c|}{\textbf{one region}} & \multicolumn{1}{c|}{\textbf{4-A}} & \multicolumn{1}{c}{\textbf{4-B}} \\ \cline{2-7} 
\multicolumn{1}{l|}{\textbf{log $\tau$}} & \multicolumn{3}{c|}{\textbf{DIC}}                                                                   & \multicolumn{3}{c}{\textbf{logML}}                                                                  \\ \hline
\textbf{-2}                              & 323553                          & 323438                          & 323629                          & -251872                         & -230926                         & -231176                        \\ \hline
\textbf{-1}                              & 311775                          & 311550                          & 311673                          & -220550                         & -199392                         & -199647                        \\ \hline
\textbf{0}                               & 303579      & 303411                          & 303580                          & -196074     & -175194  
& -175379                        \\ \hline
\textbf{1}                               & 292667      & 292435                          & 292655                          & -177810     & -156964                         & -157104                        \\ \hline
\textbf{2}                               & 281323      & 281147                          & 281169                          & -165720     & -144907                         & -144930                        \\ \hline
\textbf{3}                               & 271608      & 271590                          & 271598                          & -157987    & -137277                         & -137278                       
\end{tabular}
\caption{Comparison of the model selection criteria for the stationary Besag (one region), the \texttt{fbesag} model with 4 sub-regions (A), and the \texttt{fbesag} model with 4 sub-regions (B).}
\label{DIClogmlik:fbesag}
\end{table}

\subsection{Impact of the choice of the flexibility parameter $\sigma_\gamma$}
\begin{figure}
  \centering
  \includegraphics[scale=0.15]{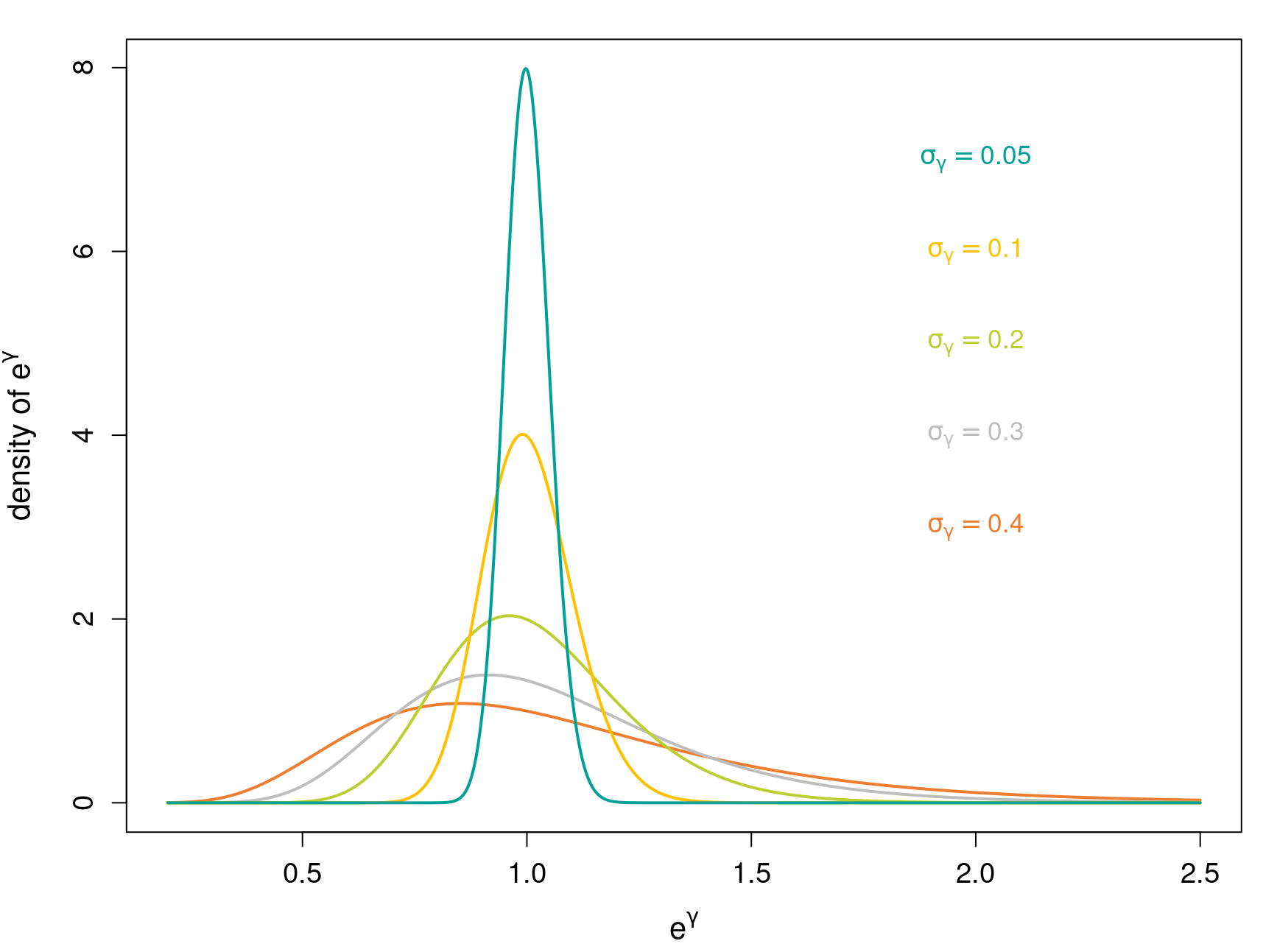}
  \caption{Density plots of \(e^{\gamma}\) for different \(\sigma_\gamma\) values such that $\gamma$ is Gaussian with mean 0 and standard deviation $\sigma_\gamma$.}
  \label{fig:sub1fig:sigmapbesag}
\end{figure}

Note that each $\tau_i = \tau e^{\gamma_i}$ for a given sub-region $i$, has a log-normal prior distribution with mean $\tau$ and variance $(e^{\sigma_{\gamma}^2} - 1)e^{2\tau + \sigma_{\gamma}^2}$, from \eqref{Jointprior}. For a given $\tau = 1$, we plot the log-normal distribution of $\tau_i$ using different values of the flexibility parameter $\sigma_\gamma = 0.05$, $0.1$, $0.2$, $0.3$, and $0.4$ in Figure \ref{fig:sub1fig:sigmapbesag}. The density ranges from flatter to peaked depending on the value of $\sigma_\gamma$. We allocate the choice of this parameter to the practitioner considering that a lower value of $\sigma_\gamma$ gives less flexibility on the precision parameters and a higher value gives more flexibility.

\begin{figure}
  \centering
  \includegraphics[scale=0.7]{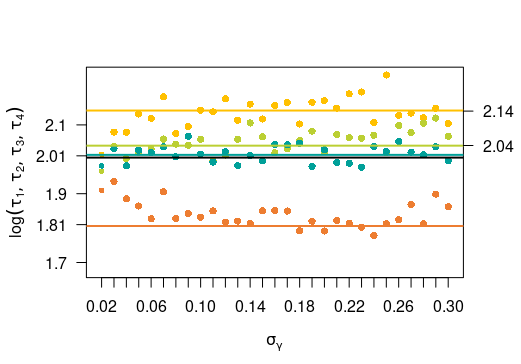}
  \caption{Local precision parameters estimated for different values of $\sigma_\gamma$. The colored lines represent the true values of the precision parameters of the four sub-regions. The black line represents the precision parameter we obtain from the stationary Besag model.}
  \label{fig:sub2fig:sigmapbesag}
\end{figure}

We  assign $\log \pmb \tau_{\text{true values}} = (2.14, 2.04, 2.01, 1.81)$ and simulate Poisson counts using the flexible Besag model with 4 sub-regions (A) as in \eqref{eq:poisson_sim}. We consider different $\sigma_\gamma$ in the range $[0.02, 0.3]$, and for each value of $\sigma_\gamma$, we fit the true model to the data. In Figure \ref{fig:sub2fig:sigmapbesag} we present the estimated local precision parameters for each value of $\sigma_\gamma$. It is clear that for small values of $\sigma_\gamma$, like $\sigma_\gamma < 0.08$ the resulting local precision parameters are estimated to be close to the precision parameter of the stationary Besag model and as such the flexible besag model contracts to the Besag model. For values $\sigma_\gamma > 0.08$, the true local precision parameters are recovered. The flexibility parameter $\sigma_\gamma$ thus impacts the model's ability to contract or deviate from the stationary model. In practice, we do not advise the use of a very small value as motivated in this section.

\vspace{0.3cm}
\subsection{Contraction of the precision parameters in the case of stationarity}
Now we simulate data using the stationary Besag model with precision parameter $\log \tau = 0.69$ for Germany using the following model 
\begin{equation}
    \pmb y \sim \text{Poisson}(\exp(\pmb 2 + \pmb \alpha_{\text{besag}})).
\end{equation}
We generate $100$ replications and for each dataset we fit four different flexible Besag model with varying number of subregions. We assign $\sigma_\gamma = 0.15$ and $\lambda = 11.51$ for the hyperprior \eqref{Jointprior}.
 We compare each fitted model with the stationary Besag model.   

\begin{table}
\centering
\resizebox{\columnwidth}{!}{%
\begin{tabular}{l|l|c|c}
\multicolumn{1}{c|}{\textbf{}} & \multicolumn{1}{c|}{\textbf{Log Precision Parameters}} & \multicolumn{1}{l|}{$\text{max}_i |0.71 - \log \tau_i|$} & \multicolumn{1}{l}{$|0.71 - \text{mean}[\log \pmb \tau]|$} \\ \hline
\textbf{Besag model}          & 0.71                                                  & 0.00                                                     & 0.00                                                   \\
\textbf{3 sub-regions}           & 0.71, 0.72, 0.70                                       & 0.01                                                     & 0.00                                                    \\
\textbf{4 sub-regions (A)}         & 0.73, 0.71, 0.72,0.72                                & 0.02                                                     & 0.01                                                    \\
\textbf{5 sub-regions}           & 0.72, 0.69, 0.70, 0.70, 0.73                           & 0.02                                                     & 0.00                                                    \\
\textbf{6 sub-regions}           & 0.72, 0.70, 0.75, 0.72, 0.69, 0.69                     & 0.04                                                     & 0.00                                                   
\end{tabular}}
\caption{Log precision values for different sub-regions. The data is simulated using the stationary Besag model.}
\label{contraction:fbesag}
\end{table}

The stationary Besag model fit results in a precision parameter estimate of $\log \tau = 0.71$. Also, the \texttt{fbesag} model with different sub-regions (see Figure \ref{fig:partgerm}) estimates log local precision parameter values very close to 0.71 for all the subregions, in each case; see Table \ref{contraction:fbesag}. This indicates the ability of the PC prior to contract the flexible Besag model to the Besag model if the data is indeed generated based on a stationary spatial field. We can thus use the \texttt{fbesag} model with the PC prior to investigate the non-stationarity of the inferred spatial field since the contraction will indicate a stationary field to be appropriate, if that is the case. Although others priors can be used for the local precision parameters, we advocate for the PC prior because of the contraction property, which cannot be guaranteed with other user-specified or convenience priors.

\section{Modeling dengue risk in Brazil using a non-stationary disease mapping model}

We revisit the model proposed by \cite{Lowe2021CombinedEO} to analyze the effects of hydrometeorological hazards on dengue risk in Brazil. To test the spatial variations in the spread of the virus in different sub-regions of Brazil, we fit dengue counts with a Poisson regression model as follows,
\begin{equation}
    \pmb y \sim \text{Poisson}(\pmb \phi, e^{\pmb \eta}), \quad \pmb \eta = \pmb 1^T \mu + \pmb \alpha + \pmb \kappa
    \label{eq:brazil}
\end{equation}
\noindent where $\pmb y$ is the observed monthly counts of dengue cases, $\pmb \phi$ is the offset, $\pmb \eta$ is the linear predictor, $\mu$ is the overall intercept, $\pmb \alpha$ is the Besag (model 0) or flexible Besag model over space, and $\pmb \kappa$ is a random walk (of order one) model over time on a torus as a temporal smoother, with precision parameter $\tau_\kappa$. 

For the flexible Besag model we consider two different sets of sub-regions, firstly we consider five sub-regions based on administrative groupings (model 1) and secondly we consider six sub-regions based on terrestrial biomes (model 2, the Pantanal region consists of only three small areas, and we combine this sub-region with the Cerrado sub-region.) (see Section \ref{sec:motivation} and Figure \ref{fig:brazilsub} for more details).
We assign the joint PC prior in \eqref{Jointprior} with $\sigma_\gamma = 0.15$ and $\lambda = 11.51$, for the flexible Besag models and the PC prior with $\lambda = 11.51$ for the Besag model. For the temporal smoother, we assign the PC prior for $\tau_\kappa$ with $u = 0.5, \alpha = 0.01$. The overall intercept is assigned a weakly informative centered Gaussian prior with low precision. We fit the models using the INLA methodology \citep{vanNiekerk2022ANA}; see code on \href{https://github.com/esmail-abdulfattah/fbesag}{github}.

\begin{figure}
    \centering
    \includegraphics[width = 8cm]{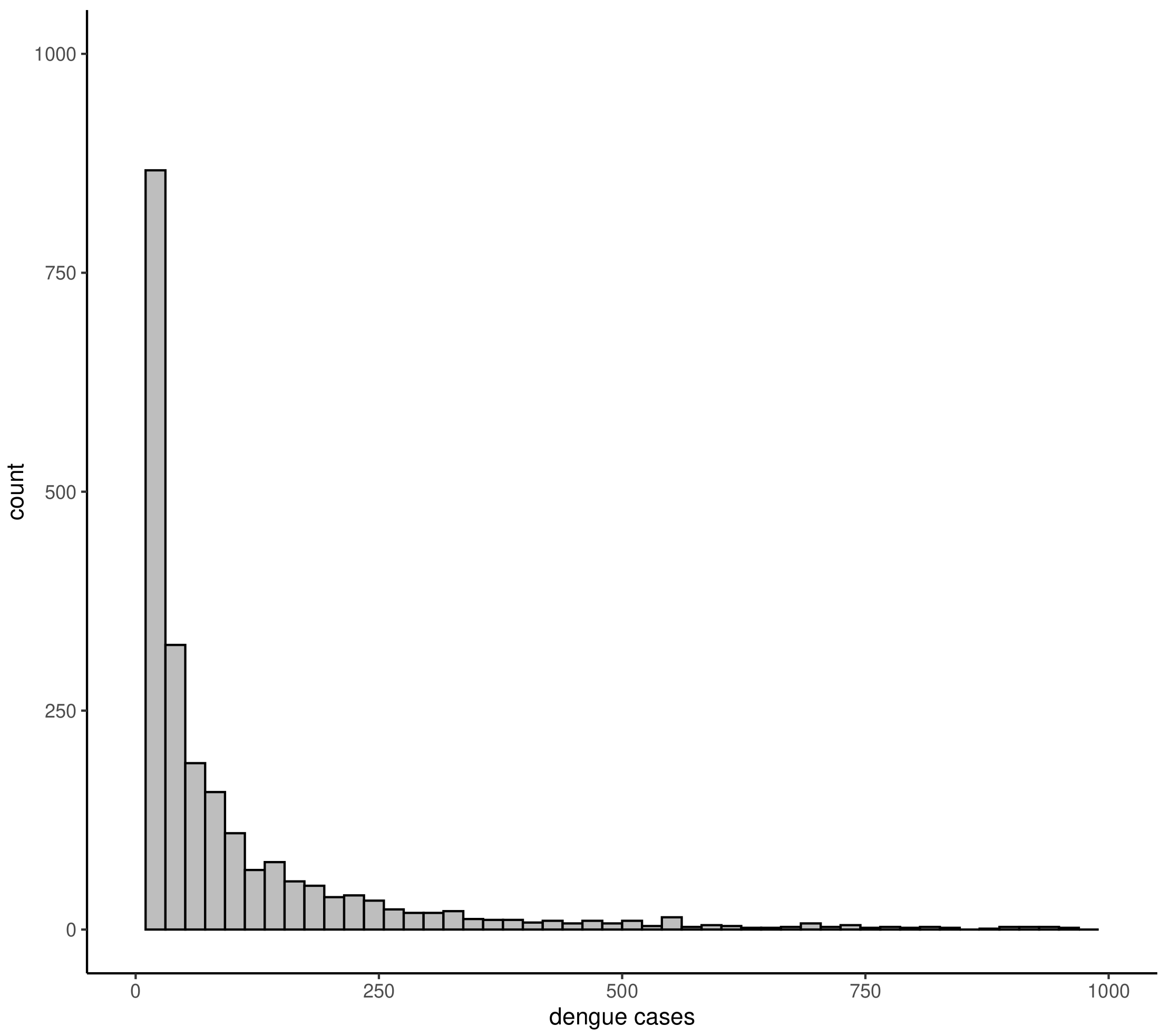}
    \includegraphics[width = 8cm]{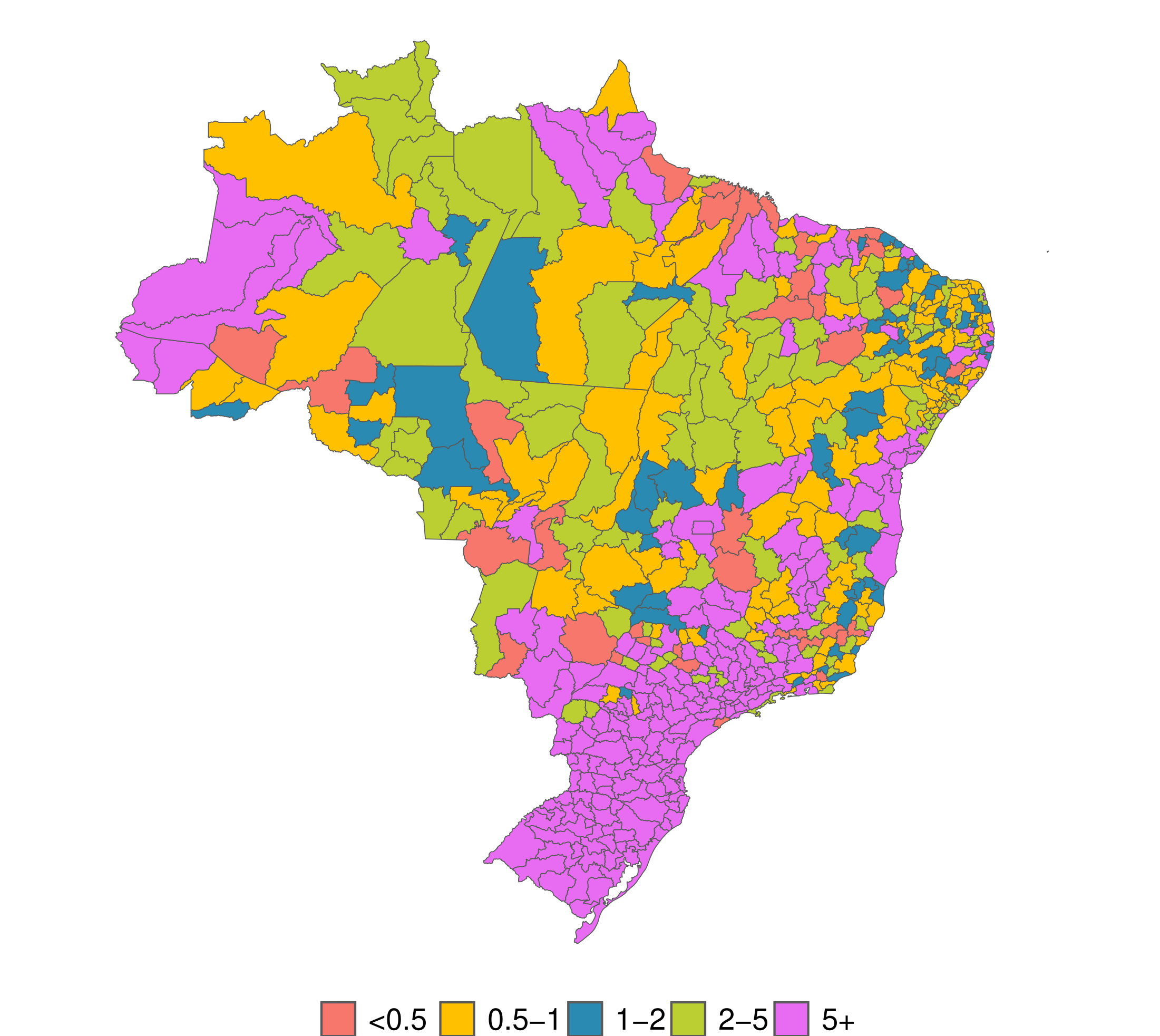}
    \caption{Observed counts (left) and SMR (right) for dengue in Brazil}
    \label{fig:brazilhist}
\end{figure}

The dataset contains 561637 cases recorded in 6696 observations (monthly data for a year across the 558 microregions). The number of cases are summarized in a histogram in Figure \ref{fig:brazilhist} (left) and it is clear that most reported numbers are low but a few are very high (60 observations were higher than 1000). In Figure \ref{fig:brazilhist} (right) we present the SMR (the ration between the observed number and the expected number if there is no difference between microregions) for dengue, which ranges from some areas with lower SMR while other areas have very high SMRs indicating that there is a significantly increased risk in those areas. Using the disease mapping model \eqref{eq:brazil} we can try to explain the increased risk of dengue using predictors and also accounting for unexplained differences in space and time.

\begin{figure}
    \centering
    \includegraphics[width=0.5\linewidth]{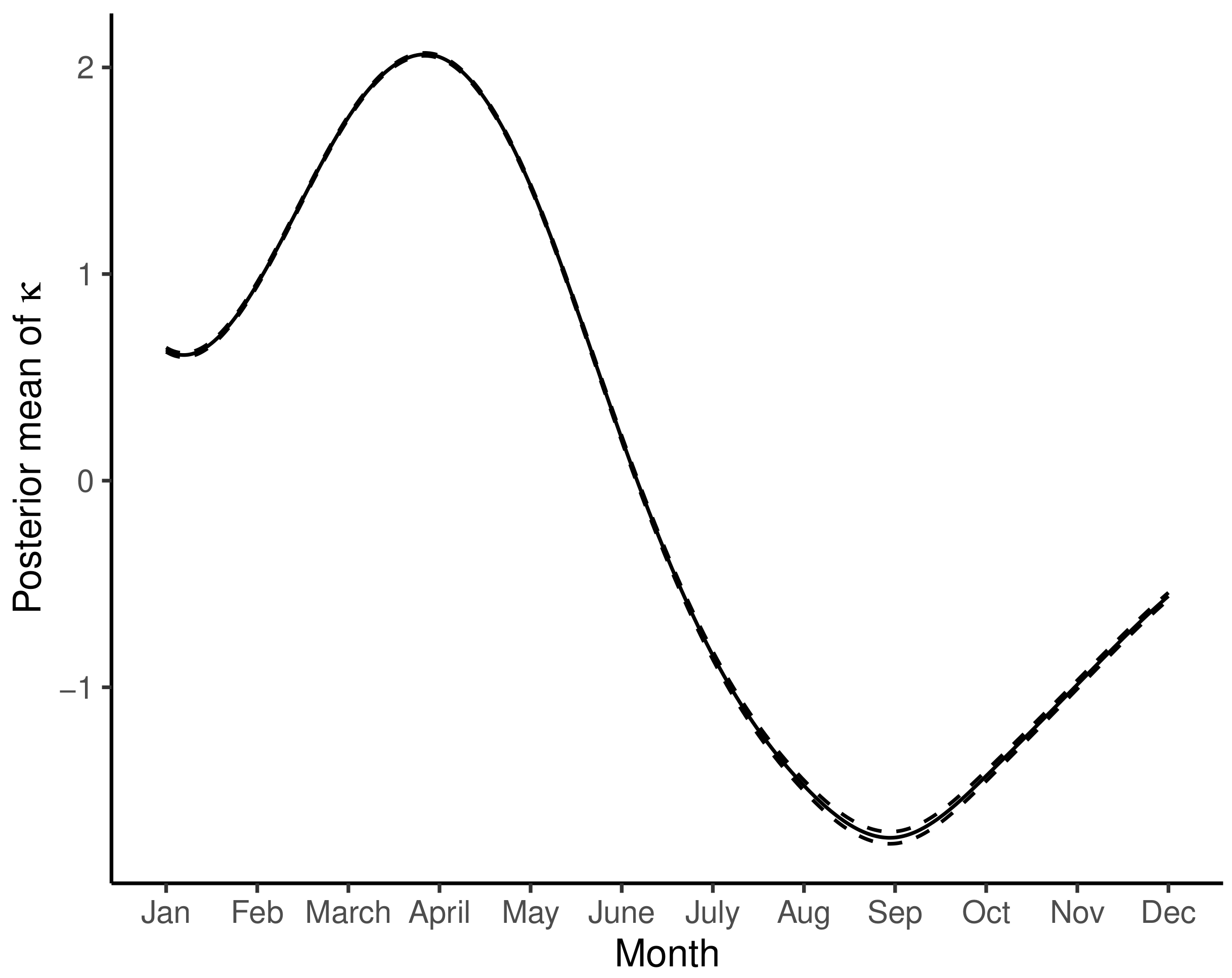}
    \caption{Temporal effect ($\kappa$) for dengue in Brazil - posterior mean (solid line) and $95\%$ credible interval (broken lines)}
    \label{fig:kappa}
\end{figure}

The inferred models are summarized in Table \ref{TAB:res}. The temporal effect and the overall intercept are very similar for the three models indicating the identifiability of the effects. The estimated temporal effect is presented in Figure \ref{fig:kappa} and the increased risk during the warmer months (January - June) is clear. Preventative controls of vectors from October onward could thus be used to lower the increased risk in the subsequent months.

\begin{table}
\centering
\renewcommand{\arraystretch}{1.3}
\begin{tabular}{c|c|c|c}
               & \textbf{Model 0} & \textbf{Model 1} & \textbf{Model 2} \\ \hline
$\mu$ & $0.737(0.708;0.766)$ & $0.741(0.712;0.77)$ & $0.735(0.706;0.765)$\\

$\pmb\tau$ & $0.303(0.267;0.342)$ & $0.327(0.249;0.421)$ & $0.387(0.309;0.479)$\\
 & & $0.414(0.342;0.495)$ & $0.297(0.237;0.366)$ \\
 & & $0.225(0.186;0.269)$ & $0.341(0.238;0.474)$\\
 & & $0.366(0.277;0.472)$ & $0.319(0.245;0.407)$\\
 & & $0.374(0.278;0.483)$ & $0.276(0.229;0.327)$\\
$\tau_\kappa$ & $1.972(0.961;3.386)$ & $1.945(0.985;3.437)$ & $1.946(0.975;3.383)$\\ \hline
\textbf{DIC}   & -73259           & \textbf{-73163}          & -73186          \\ \hline
\textbf{LCVS}  & 16.551            & 16.551             & 16.549           \\ \hline
\textbf{logML} & -118855        & \textbf{-118603}        & -118612       
\end{tabular}
\caption{Comparison of models 0, 1, and 2 using different criteria.}
\label{TAB:res}
\end{table}

 We compare models 1 and 2 with model 0 using the deviance information criterion (DIC), the log cross-validation score (LCVS) \citep{Liu2022LeavegroupoutCF}, and the log marginal likelihood (logML). We summarize the results in Table \ref{TAB:res}. 

Both models 1 and 2 show improved goodness of fit based on the three criteria, when compared to model 0. The inferred precision parameters in model 1 are $\pmb \tau_1 = (0.327, 0.414, 0.225, 0.366, 0.374) $ for the North, Northeast, Central-West, Southeast, and South sub-regions, respectively. In model 2, the inferred precision parameters are  $\pmb \tau_2 = (0.387, 0.297, 0.341, 0.319, 0.276)$. Model 0 gives $ \tau_0 = 0.303$.  

\begin{figure}
\centering
\begin{subfigure}[b]{0.5\textwidth}
    \centering
  \includegraphics[width=\linewidth]{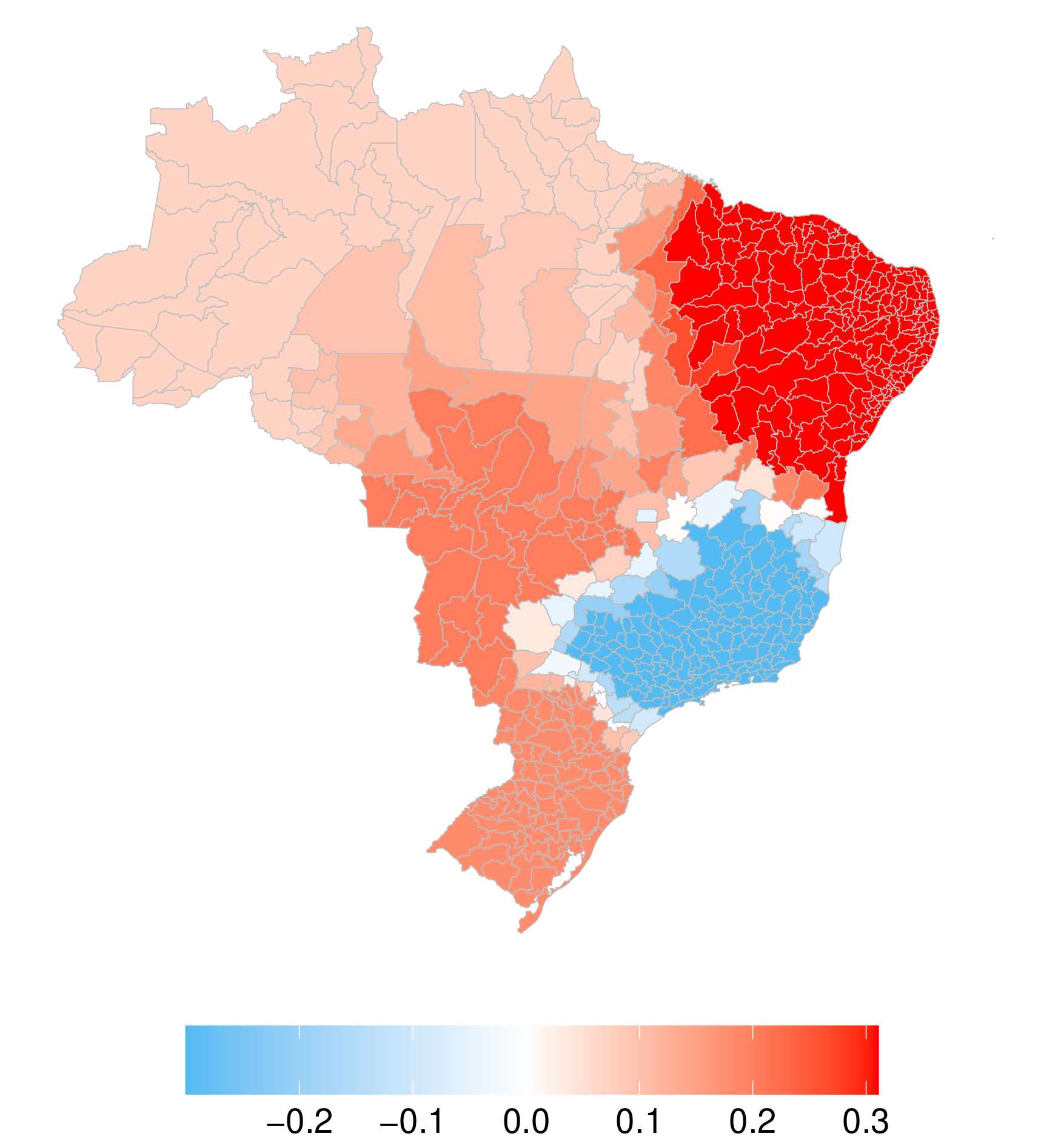}
   \caption{$\log\pmb\tau$}
\end{subfigure}%
\begin{subfigure}[b]{0.5\textwidth}
    \centering
  \includegraphics[width=\linewidth]{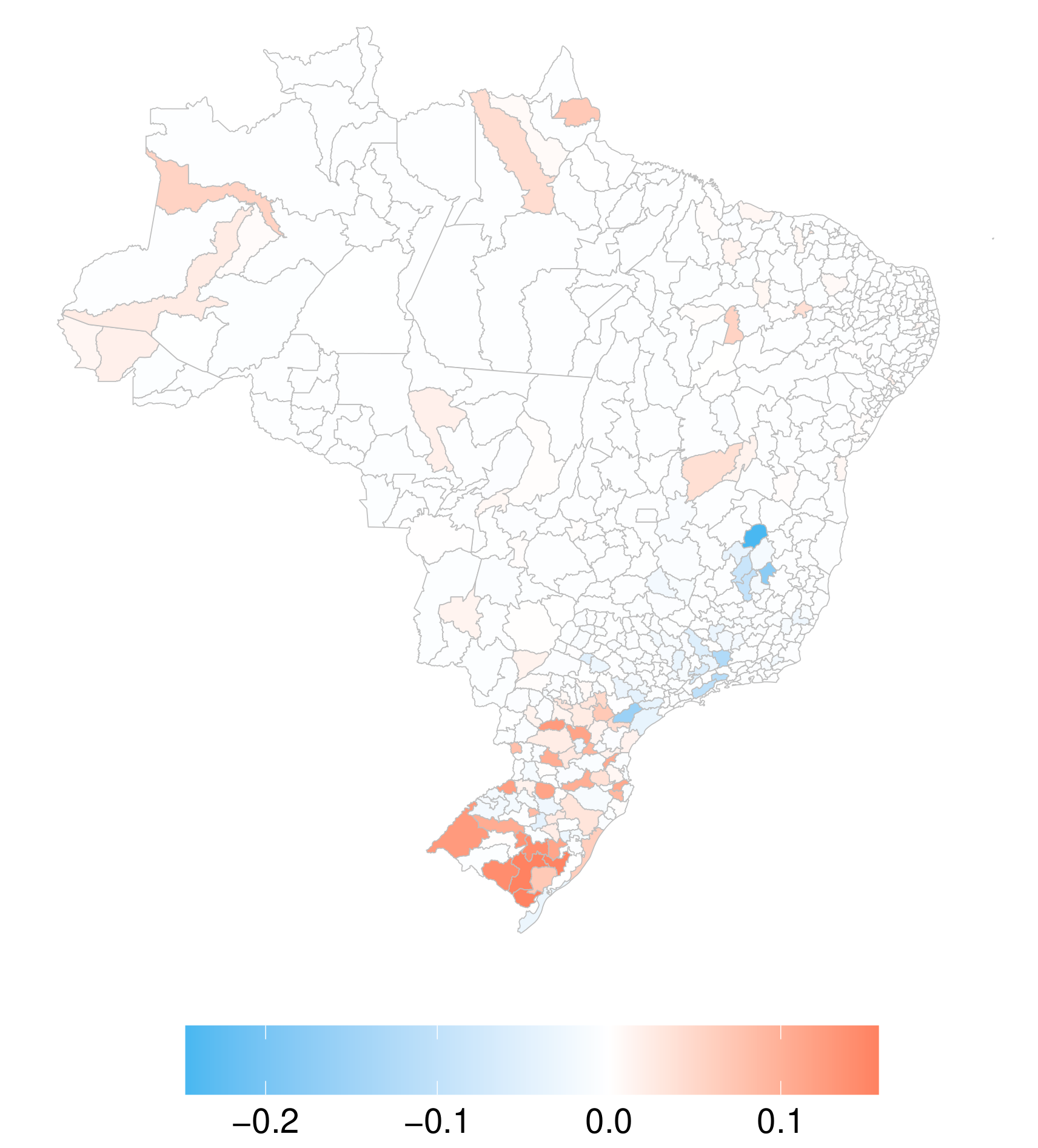}
   \caption{Posterior mean of $\pmb\alpha$}
\end{subfigure}

\begin{subfigure}[b]{0.5\textwidth}

  \includegraphics[width=\linewidth]{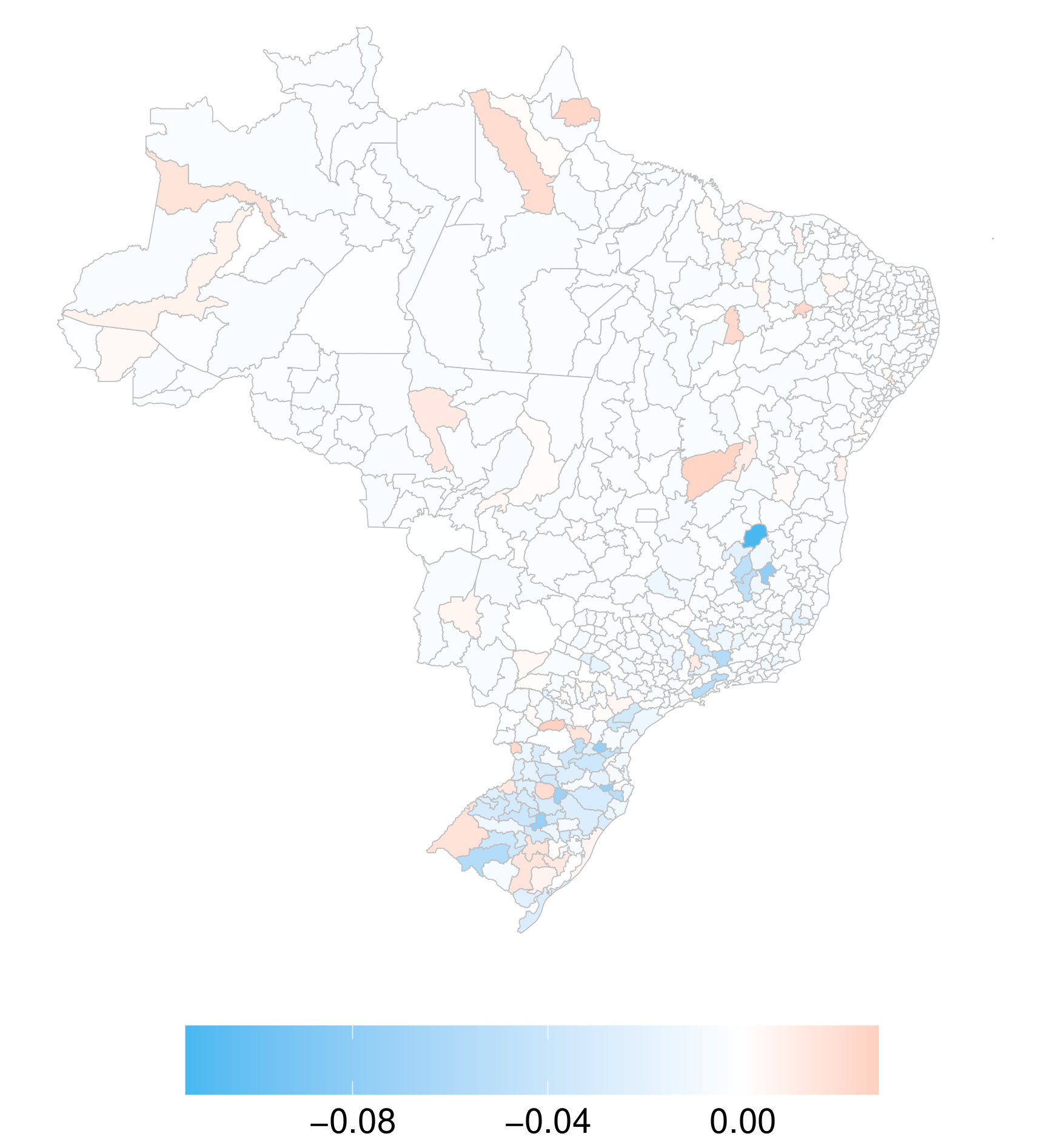}
  \caption{$97.5^{\text{th}}$ posterior percentile of $\pmb\alpha$}
\end{subfigure}%
\begin{subfigure}[b]{0.5\textwidth}

  \includegraphics[width=\linewidth]{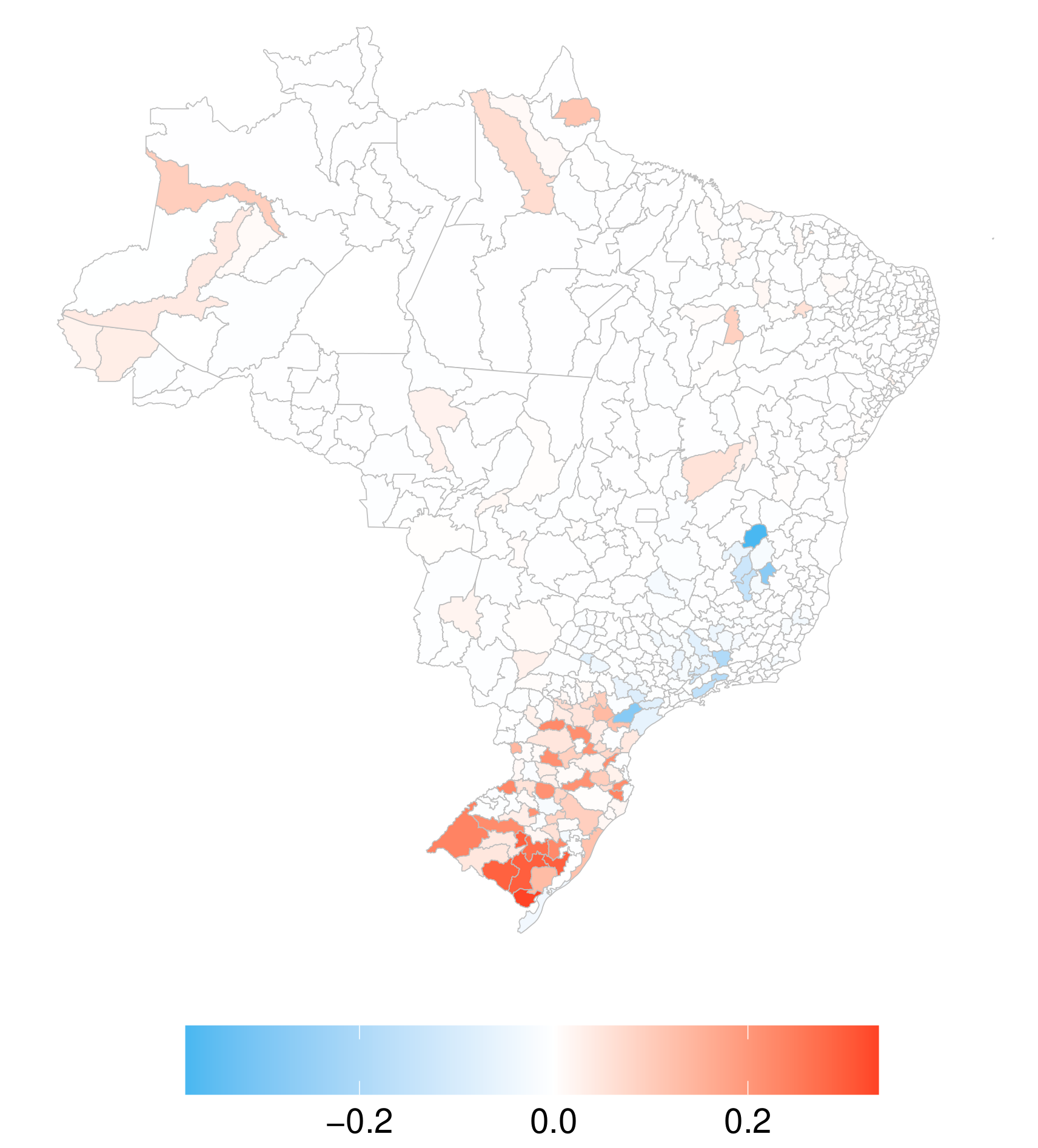}
   \caption{$2.5^{\text{th}}$ posterior percentile of $\pmb\alpha$}
\end{subfigure}
\caption{Difference between the \texttt{fbesag} (model 1) and Besag (model 0) models }
\label{fig:partitions_brazil5}
\end{figure}

In Figure \ref{fig:partitions_brazil5}, we plot the difference between the flexible Besag model 1 and the stationary Besag model 0 as characterized by the difference in the posterior mean and tail percentiles of $\pmb\alpha$ and the posterior mean of $\log\pmb\tau$. The local precision parameters are quite different from the stationary precision parameter for most regions (the non-white areas). The sub-region of the Northeast exhibits the weakest spatial dependence structure, while the Southeast sub-region displays the strongest spatial dependence structure from Figure \ref{fig:partitions_brazil5} (a). The non-stationarity of the spatial field also changes the posterior mean of the Gaussian field as shown in Figure \ref{fig:partitions_brazil5} (b). Note that in the Southeast sub-region, where a stronger spatial dependence structure is estimated, some microregions have a much lower estimated risk (blue regions) than under the stationary model. In the South sub-region where the local precision parameter is higher than the precision parameter of the stationary model, various microregions have a higher estimated risk (red regions). Considering Figures \ref{fig:partitions_brazil5} (c) and (d), we note that for most microregions the $95\%$ credible intervals are narrower than those estimated under the stationary model. 
 
\begin{figure}
\centering
\begin{subfigure}[b]{0.5\textwidth}

  \includegraphics[width=\linewidth]{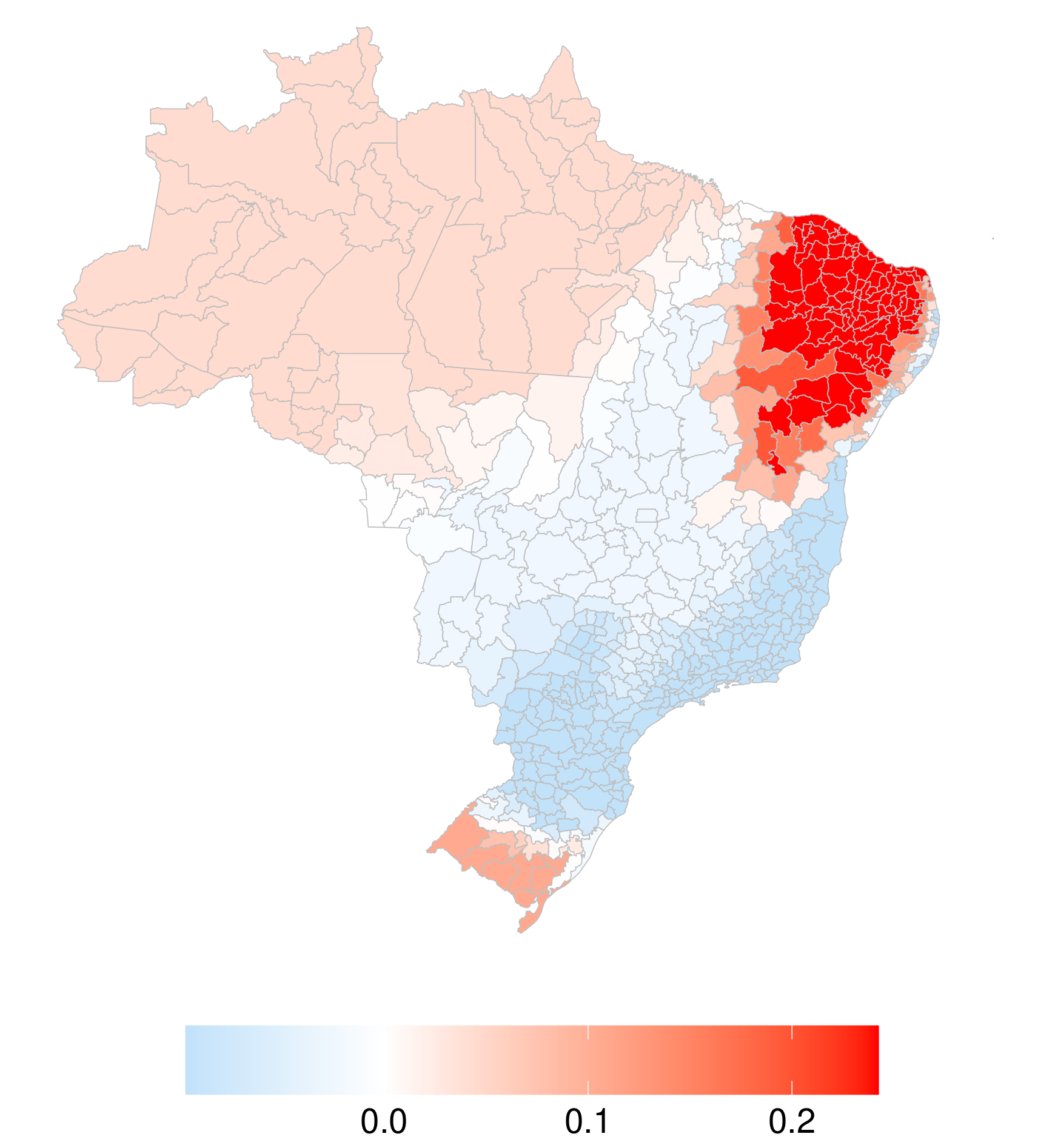}
   \caption{$\log\pmb\tau$}
\end{subfigure}%
\begin{subfigure}[b]{0.5\textwidth}

  \includegraphics[width=\linewidth]{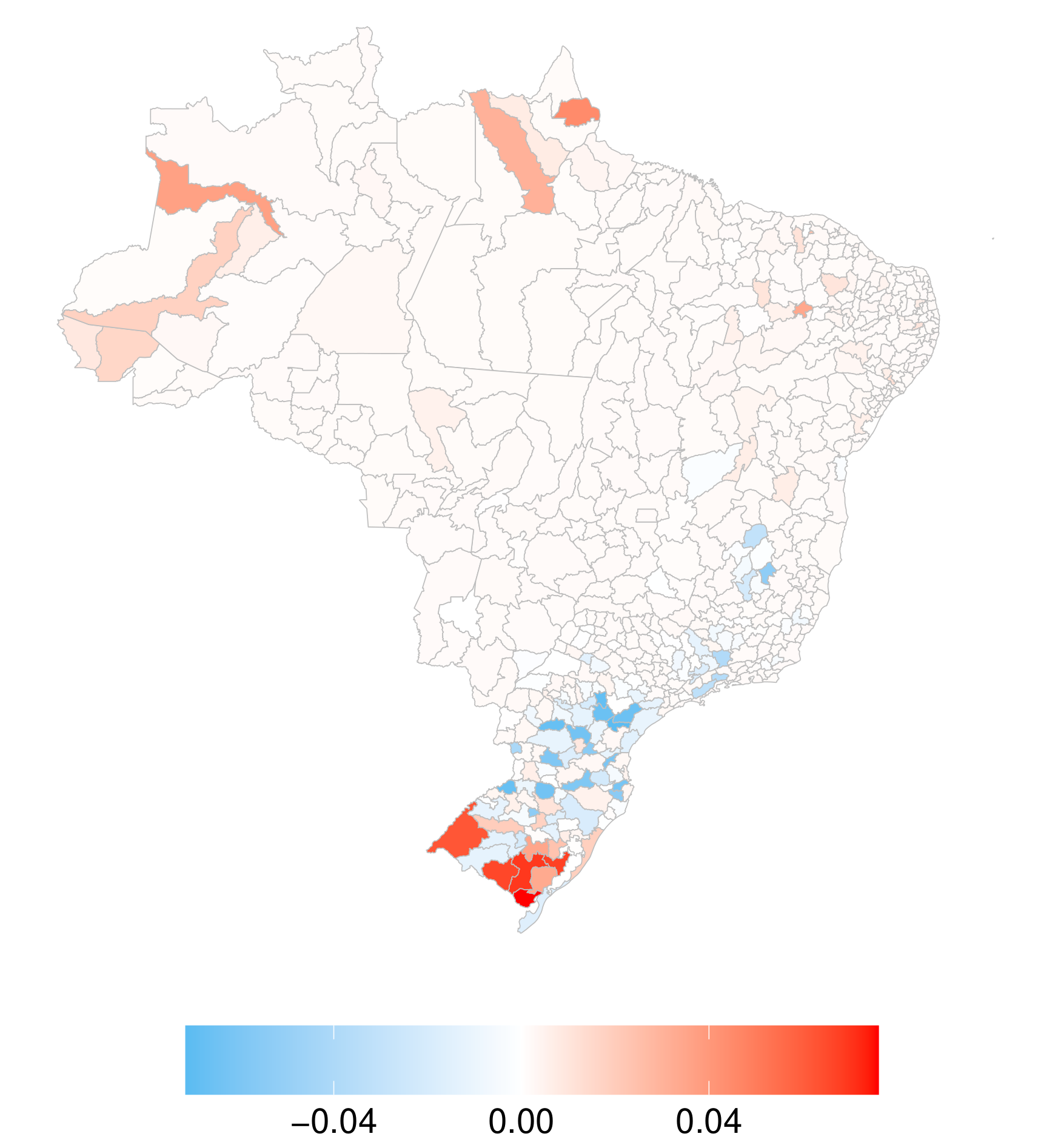}
   \caption{Posterior mean of $\pmb\alpha$}
\end{subfigure}

\begin{subfigure}[b]{0.5\textwidth}

  \includegraphics[width=\linewidth]{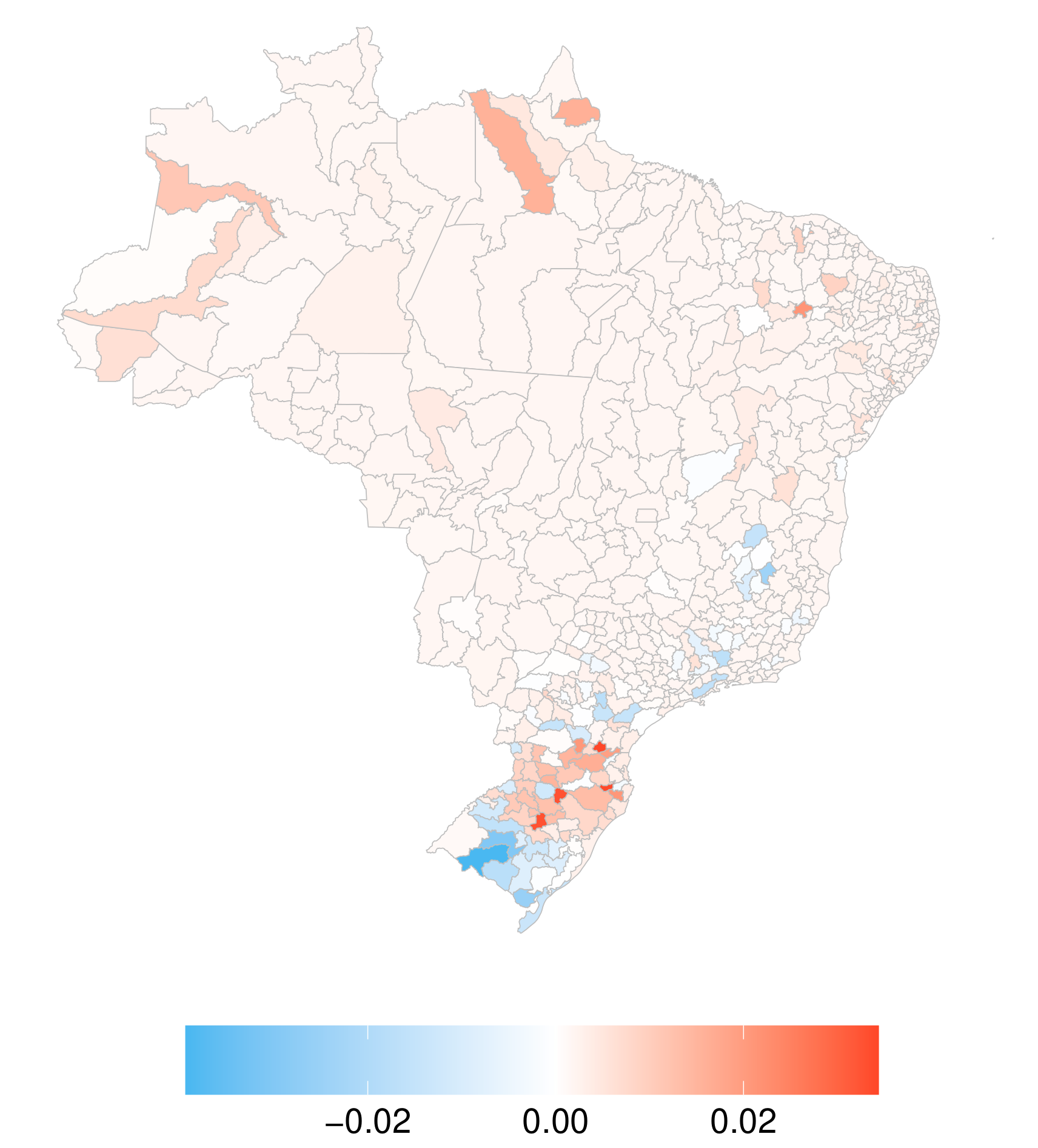}
  \caption{$97.5^{\text{th}}$ posterior percentile of $\pmb\alpha$}
\end{subfigure}%
\begin{subfigure}[b]{0.5\textwidth}

  \includegraphics[width=\linewidth]{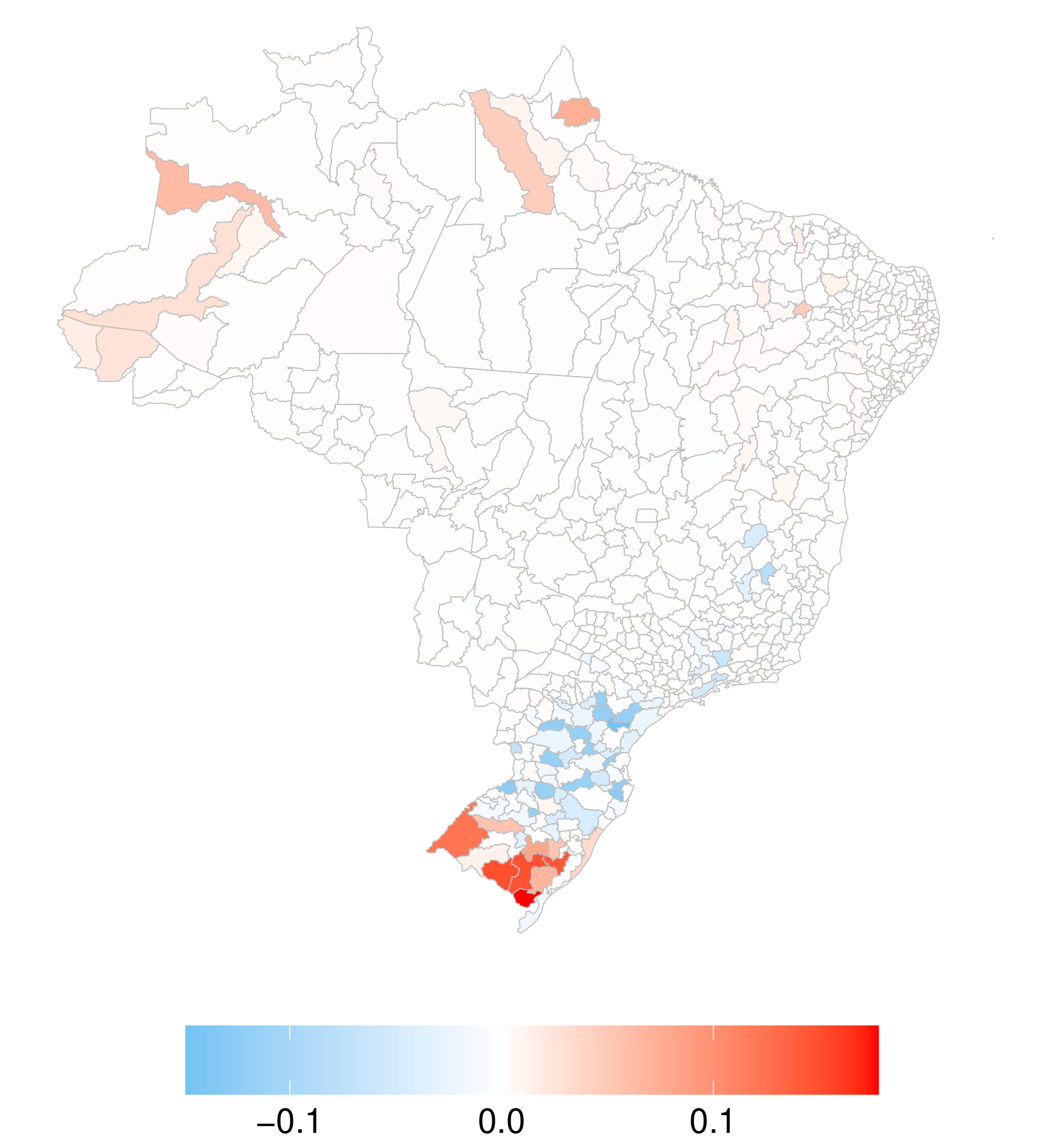}
   \caption{$2.5^{\text{th}}$ posterior percentile of $\pmb\alpha$}
\end{subfigure}
\caption{Difference between the \texttt{fbesag} (model 2) and Besag (model 0) models}
\label{fig:partitions_brazil6}
\end{figure}

  Similar interpretations can be made for Figure \ref{fig:partitions_brazil6}, while noting that the differences are not as large as those between model 1 and model 0. Interestingly, both non-stationary models highlight similar regions where the risk of dengue is increased (red) or decreased (blue) which would have gone unnoticed with the stationary model.  
When considering the DIC, LCVS and the logML, together with the lower uncertainty and substantial difference in the posterior mean of the spatial field, model 1 has a larger impact on risk estimation than model 2, and can therefore be preferred.   

\section{Conclusions}\label{secconclusions} 

Stationarity of spatial fields are often assumed due to the lack of an alternative model. In disease mapping, the stationary Besag model has been, and still is, used very often to model spatial variation of a disease as a means to understand unexplained variability in the risk of a disease, and subsequently inform interventions and policies. In this work we propose a non-stationary counterpart of the stationary Besag model, that can account for different spatial intensities over a study region. This proposal can discover hidden spatial patterns that may help policymakers, public health sectors, and epidemiologists, amongst others in several ways. 

The proposed flexible Besag model is capable of determining whether there is uniform spatial variation across all sub-regions or not, due to the use of a joint penalizing complexity prior that can contract the non-stationary model to the stationary model in the case of stationarity. The practitioner, thus can use the proposal as an investigative tool to evaluate the need for a non-stationary spatial field, without the risk of overfitting, as with many other flexible models. 

In the case of dengue in Brazil, the stationary assumption of the spatial field is not realistic due to the large study region and various substantial differences in sub-regions. We showed that the estimated risk profiles for the microregions are negatively impacted by the assumption of the stationary model, thus the non-stationary model is necessary for more accurate and precise risk predictions. 

Naturally, the proposed model can be extended to a non-stationary BYM (and BYM2) model \citep{Riebler2016AnIB} based on sub-regions. Further research is needed to develop models that incorporate space-time interactions based on sub-regions. The \texttt{fbesag} model could be explored to fit the spatiotemporal models with the interaction types \rom{1}, \rom{2}, \rom{3}, and \rom{4} that are presented in \citep{type1234}. The new model is implemented in the accompanying R package \textbf{\texttt{fbesag}} available on github at \textcolor{blue}{esmail-abdulfattah/fbesag}; see Appendix \ref{appendix::installfbesag}, that can be used to incorporate the \texttt{fbesag} model in the R library \texttt{INLA} for convenient use by practitioners. Based on the details presented in this paper, the \texttt{fbesag} model can be easily implemented as a latent effect in various other user-preferred inferential frameworks.

We can extend this idea further, to domains other than space. We can use the model to account for non-stationarity in time by assigning different precision parameters for different seasons; For instance in the Brazil data, we could assign $\tau_1$ for months $1,2,3,4$, $\tau_2$ for the months $5,6,7,8$ and $\tau_3$ for the months $9,10,11,12$, or we use 12 $\tau$'s for the 12 months over multiple years. By allowing different precision parameters for each season, the model can account for seasonal variations in disease incidence or mortality. For example, certain diseases may exhibit different patterns of occurrence during different seasons. Mosquito-borne diseases like dengue or malaria might be more prevalent during the rainy season, while respiratory illnesses could be more common during colder months.  

We believe that this work is important for epidemiologists, data scientists and statisticians working with disease mapping and other stationary latent effects, since the proposal allows an intuitive way to account for non-stationarity in well-known stationary models. It can be used as a basis for various future developments of non-stationary Gaussian effects such as temporal model, spatial models and stochastic spline models.

\newpage
\appendix
\section*{Appendices}
\addcontentsline{toc}{section}{Appendices}
\renewcommand{\thesubsection}{\Alph{subsection}}

\subsection{Joint PC prior for precision parameters} \label{appendix::pbesagPC}

We follow the penalized complexity prior (PC) framework to derive a joint prior for the mixing precision parameters that give more flexibility to the level of smoothing over space by partitioning the small areas into groups and preventing over-fitting based on four principles, summarized next.

Given $p$ partitions, let $\pi_1$ denote the density of a partitioned Besag model $\pmb x \in \mathbb{R}^{n}$ with  precision parameters $\pmb \tau = (\tau_{1}, \ldots, \tau_{p})$ based on equation \ref{eq::besagtype1}, such that $\pmb \tau = \tau e^{\pmb \gamma}$. This model can be seen as a flexible extension of a based model with density $\pi_0$ and $\pmb \gamma = \pmb 0$ (i.e. absence of clusters). The four principles are 
\begin{enumerate}
    \item Parsimony: The joint prior for $\pmb \gamma = (\gamma_1, \ldots \gamma_p)$ should give proper shrinkage to $\pmb \gamma = \pmb 0$ and decay with the increasing complexity of $\pi_1$ so that the simplest model is favored unless there is evidence for a more flexible one.
    \item The increased complexity of $\pi_1$ with respect to $\pi_0$ is measured using the Kullback-Leibler divergence KLD \citep{Kullback1951OnIA},
    \begin{equation}
        \text{KLD}(\pi_1 ||\pi_0) = \displaystyle \int \pi_1 (\pmb \tau) \log \Bigg ( \frac{\pi_1 (\pmb \tau) }{\pi_0 (\pmb \tau)} \Bigg) d \pmb \tau.
    \end{equation}
    For ease of interpretation, the \text{KLD} is transformed to a unidirectional distance measure 
    \begin{equation}
        \text{d}(\pmb \tau) = \text{d}(\pi_1 || \pi_0) = \sqrt{2 \text{KLD}(\pi_1||\pi_0)} 
    \end{equation}
    that can be interpreted as the distance from the flexible model $\pi_1$ to the base model $\pi_0$.

    Let $ \mathcal{N}(\pmb \mu, \pmb \Sigma) $ denote a multivariate normal distribution with dimension $n$. The Kullback-Leibler divergence from $ \mathcal{N}_1(\pmb \mu_1, \pmb \Sigma_1) $ to $ \mathcal{N}_0(\pmb \mu_0, \pmb \Sigma_0) $ for a latent field $\pmb x$ is,
\begin{equation}
\begin{split}
\text{KLD}(\mathcal{N}_1 || \mathcal{N}_0) & = \displaystyle \int (2\pi)^{-n/2} |(\pmb \Sigma_1)|^{-1/2} e^{-\frac{1}{2}(\pmb x - \pmb \mu_1)^{\!{\mathsf {T}}} \pmb \Sigma_1^{-1} (\pmb x - \pmb \mu_1) } \\ & ~~~ \log \Bigg(\frac{ |(\pmb \Sigma_1)| ^{-1/2} e^{-\frac{1}{2}(\pmb x - \pmb \mu_1)^{\!{\mathsf {T}}} \pmb \Sigma_1^{-1} (\pmb x - \pmb \mu_1) }}{ |(\pmb \Sigma_0)| ^{-1/2} e^{-\frac{1}{2}(\pmb x - \pmb \mu_0)^{\!{\mathsf {T}}} \pmb \Sigma_0^{-1} (\pmb x - \pmb \mu_0) }} \Bigg) d\pmb x \\ & =  \dfrac{1}{2} \Big\{ - \log \frac{|\pmb \Sigma_1|}{|\pmb \Sigma_0|} + tr (\pmb \Sigma_0^{-1}\pmb \Sigma_1 ) + (\pmb \mu_0 - \pmb \mu_1)^T \pmb \Sigma_0^{-1} (\pmb \mu_0 - \pmb \mu_1) - n \Big\}.
\end{split}
\end{equation}

Define the general form of the precision matrix of the flexible Besag model with $\epsilon \pmb I$ added to the diagonal and as a function of $\pmb \tau = (\tau_1, \ldots, \tau_p)$,
\begin{equation}
    \pmb \Sigma^{-1}_1 = \pmb M_1 (\pmb \tau)
\end{equation}
where $\pmb M_1 = \pmb R (\pmb \tau) + \epsilon_1 \pmb I$. This formulation can be extended to derive a partitioned bym2 model \citep{Leroux2000EstimationOD}. The base model to find the PC-prior for the precision parameter in Besag model is,
\begin{equation}
    \pmb \Sigma^{-1}_0 = \tau \pmb R, \quad \tau \to \infty,
\end{equation}
and the base model to find the PC-prior of $\epsilon$, 
\begin{equation}
    \Sigma^{-1}_0 = \pmb M_0 (\pmb \tau), \quad \epsilon \to 0, ,
\end{equation}
where $\pmb M_0 = \pmb R (\pmb \tau) + \epsilon_0 \pmb I$ and $\pmb \tau = (\tau, \ldots, \tau)$ is constant.

Let $\pmb U \Lambda \pmb U^{-1} = \sum_i \lambda_i u_i u_i^T$ be the eigen decomposition of $\pmb R$ where $\pmb u_i$ is the $i^{th}$ vector of $\pmb U$ containing the $i^{th}$ eigenvector and $\Lambda$ is a diagonal matrix with $\Lambda_{i,i} = \lambda_i$, the $i^{th}$ eigenvalue.
    \item The PC prior is defined as an exponential distribution on the distance, 
    \begin{equation}
        \pi(\text{d}(\pmb \tau)) = \lambda \exp(- \lambda \text{d}(\pmb \tau)),
    \end{equation}
    with rate $\lambda > 0$. \cite{Simpson2014PenalisingMC} computed this distance and derived the PC prior for $\theta = \log \tau$ for non-partitioned Besag model,
    \begin{equation}
        \pi(\theta) = \frac{\lambda}{2}e^{-\theta/2 -\lambda e^{-\theta/2}}
    \end{equation}
    Similarly, we get the distance,  
    \begin{equation}
        d(\epsilon_1) = \sqrt{\epsilon_1^2 \sum_i \lambda_i} \sim \exp(\lambda),
    \end{equation}
    to find the PC-prior of $\epsilon$, and the distribution of $\theta = \log \epsilon$,
    \begin{equation}
        \pi(\theta) = \lambda e^{-\lambda e^{\theta}} e^{\theta}.
    \end{equation}
    Our purpose is to find the PC-prior for,
    \begin{equation}
        \pmb \theta = \log \pmb \tau = \log \tau + \pmb \gamma = \tilde{\theta} + \pmb \gamma.
    \end{equation}
    where $\tilde{\theta} = p^{-1} \sum_i \theta_i = \overline{\theta}$.
    Based on \cite{Fuglstad2018ConstructingPT}, we assign a multivariate Gaussian prior for $\pmb \gamma = (\gamma_1, \ldots, \gamma_p)$,
    \begin{equation}
        \pmb \gamma \sim \mathcal{N}(\pmb 0, \sigma_\gamma^2 \pmb I_p) \text{ and }  \pmb \gamma | \pmb 1^T \pmb \gamma = 0 \sim \mathcal{N} \Big(\pmb 0, \tilde{\pmb \Sigma}^{-1} = \sigma_\gamma^2 ( \pmb I_p - p^{-1} \pmb 1_{p \times p}) \Big).
    \end{equation}
    and we have a PC-prior for $\overline{\theta}$,
    \begin{equation}
        \pi(\overline{\theta}) = \frac{\lambda}{2}e^{-\overline{\theta}/2 -\lambda e^{-\overline{\theta}/2}}
    \end{equation}
    By change of variable transformation,
    \begin{equation}
    \begin{split}
         \pi(\pmb \theta) &= \pi(\overline{\theta}) \pi(\pmb \gamma| \pmb 1^T \pmb \gamma = 0) |\text{J}|\\
         &=  \frac{\lambda |\text{J}|^{1/2}}{2 (\sigma_\gamma^2 2\pi)^{p/2}}  e^{-\frac{1}{2} (\pmb \theta- \pmb 1 \overline{\theta})^T \tilde{\pmb \Sigma}^{-1} (\pmb \theta- \overline{\theta} \pmb 1) - \overline{\theta}/2 - \lambda e^{-\overline{\theta}/2}}
    \end{split}
    \end{equation}
    where $\text{J}$ is the Jacobin. The log joint distribution,
    \begin{equation}
        \begin{split}
             \log \pi(\pmb \theta) &= \log(\lambda) - \log(2 (\sigma_\gamma^2 2\pi)^{P/2}) -\frac{1}{2} (\pmb \theta- \pmb 1 \overline{\theta})^T \tilde{\pmb \Sigma}^{-1} (\pmb \theta- \overline{\theta} \pmb 1) - \overline{\theta}/2 - \lambda e^{-\overline{\theta}/2}.
        \end{split}
    \end{equation}
    \item The parameters $\lambda$ and $\sigma_\gamma$ can be selected by the user based on his prior knowledge of $\pmb \tau$ (or an interpretable transformation of it such as the standard deviation).
\end{enumerate}

\subsection{fbesag Installation and Example} \label{appendix::installfbesag}

\subsubsection{Installation}
\begin{lstlisting}[style=Rstyle]
    library("devtools")
    install_github("esmail-abdulfattah/fbesag")
\end{lstlisting}

\subsubsection{Example}

We present below how to form the formula for \texttt{inla()} function in \texttt{R-INLA} package for Besag and flexible Besag models.

\vspace{0.4cm}
\begin{lstlisting}[style=Rstyle]
    library("INLA")
    library("fbesag")
    
    graph = #adjaceny spatial structure matrix of the region
    id_regions = #id for the sub-regions 
    sd_gamma = #flexibility parameter (say 0.15)
    param = #the parameters of the pc prior (u, alpha)
    id = #id of the random effects

    #Besag
    formula1 = y ~ 1 + f(id, model="generic", Cmatrix = inla.as.sparse(graph),
                         constr = TRUE, rankdef=1)
    
    #flexible Besag
    cmodel_fbesag <- get_fbesag(graph, id_regions, sd_gamma, 
                                param = list(p1 = 1, p2 = 1e-5))
    formula2 = y ~ 1 + f(id, model = cmodel_fbesag, constr= TRUE, rankdef=1)

\end{lstlisting}
\end{document}